\documentclass[titlepage,12pt]{article}

\usepackage{amsthm,amsmath,amsfonts,amssymb}
\usepackage{mathpazo}
\usepackage[onehalfspacing]{setspace}
\usepackage[margin=1in]{geometry}

\usepackage{graphicx}

\usepackage[dvipsnames]{xcolor}
 \usepackage {ulem}

\usepackage{enumerate}

\usepackage{mathdots}
\usepackage{yhmath}
\usepackage{cancel}
\usepackage{color}
\usepackage{array}
\usepackage{multirow}

\usepackage{tabularx}
\usepackage{extarrows}
\usepackage{booktabs}
\usepackage{soul}

\usepackage{hyperref}
\usepackage{multimedia}
\usepackage{rotating}

\usepackage{tikz}
\usetikzlibrary{fadings}
\usetikzlibrary{patterns}
\usetikzlibrary{shadows.blur}
\usetikzlibrary{shapes}

\usepackage{comment}

\usepackage[english]{babel}
\usepackage{hyperref}

\addto\extrasenglish{%
}

\usepackage{url}

\usepackage{natbib}
\usepackage{enumitem}
\usepackage{subcaption}

\newtheorem{theorem}{Theorem}
\newtheorem*{theorem*}{Theorem}
\newtheorem{lemma}{Lemma}
\newtheorem{proposition}{Proposition}

\newtheorem{definition}{Definition}

\theoremstyle{definition}
\newtheorem{example}{Example}

\theoremstyle{remark}
\newtheorem{remark}{Remark}

\theoremstyle{plain}

\def\mcR{\mathcal{R}}

\newcounter{resetdummycounter}
\newcommand{\resetcounterlist}[1]{%
  \renewcommand*{\do}[1]{\counterwithin*{##1}{resetdummycounter}}%
  \docsvlist{#1}}
\newcommand{\resetcounters}{\stepcounter{resetdummycounter}}
\resetcounterlist{definition, theorem, proposition, corollary}

\begin{document}

\title{Optimal Incentive-Compatible Assignments with Endogenous School Priorities
\thanks{We would like to thank  Vincent Meisner, Humberto Moreira, Francisco Silva, Paulo Somaini, and Maren Vairo  for their useful comments and suggestions. 
We are also indebted to participants at the 9º Coloquio Uruguayo de Matemática. Financial support from ANII, FCE\_1\_2025\_1\_186685, is gratefully acknowledged.}}
\author{Federico Echenique\textsc{\thanks{%
Department of Economics, University of California, Berkeley (email: \url{fede@econ.berkeley.edu)}. }} \ and \ Juan S. Pereyra\textsc{\thanks{%
Department of Economics, Universidad de Montevideo, Montevideo, Uruguay
(email: \url{jspereyra@um.edu.uy}).}}}

\maketitle

\begin{abstract}
We study whether a planner can favor students at schools they rank more highly while preserving truthful reporting. The standard implementation of the deferred acceptance (DA) algorithm is typically not the solution. With interior cutoffs, DA with multiple tie-breaking (DA-MTB) solves the planner's optimization problem if and only if the planner’s payoff from a school assignment is independent of the students' preferences. This is typically not the case, and strict incentive compatibility under DA-MTB leaves room for improvement via preference-dependent priority adjustments. We characterize optimal rank-dependent cutoff adjustments in a two-school model and show that, in a many-school environment, near-optimal approximately incentive-compatible assignments can be chosen with small support (so each type of student may only attend a small number of schools).
\end{abstract}

\clearpage
\thispagestyle{empty}
\tableofcontents
\clearpage \setcounter{page}{1}

\section{Introduction}\label{sec:intro}

We study the design of school choice programs when the planner's objective for a student depends on the student's reported preferences. School choice programs worldwide often prioritize admission for a student at a school when they rank that school highly. Programs implement this favoritism through rank-dependent priority scores, first-choice eligibility requirements, and the design of the assignment procedure. A district's motivation can be to promote student welfare or improve match quality, treating a high rank as evidence of the fit between a student and the school, or even addressing distributional considerations. In pursuit of this objective, the resulting school choice mechanisms often induce students to misreport their preferences and weaken the information that the procedure relies on. \textit{We study how a school district can optimally assign students when it wants to favor highly ranked matches while preserving incentives for truthful reporting.} We treat school priorities as design choices, and we examine how the resulting optimal assignments relate to deferred acceptance (DA), the canonical school choice algorithm.

There are many examples of centralized school choice programs that prioritize matches between students and the schools that they rank highly. In the French-speaking region of Belgium, the centralized secondary-school system runs a student-proposing DA; however, baseline priorities (e.g., sibling, special needs) apply only at a student's first-choice school, and the tie-breaking index weights a school more when the student ranks it higher \citep{cantillon2015belgium}. Taiwan provides an even more striking example: under the Taiwanese mechanism (studied in the paper by \citet{dur2022deduction}), points are directly deducted from a student's priority score when a school appears lower in their submitted ranking. Similar policies appear elsewhere. Trinity College at the University of Toronto considers only applicants who rank it first; the School District of Philadelphia guarantees continued enrollment at some schools only for those students who rank their current school first; and New York City's high-school admissions system awarded priority to students who show interest by attending information sessions or open houses. See Table~\ref{tab:preference-dependent-priorities}. These and other examples illustrate a broader institutional objective: schools often wish to favor applicants who show a strong desire to attend.

\begin{table}[htbp]
\centering
\small
\caption{Examples of admission rules in which expressed preferences affect priority or eligibility}
\label{tab:preference-dependent-priorities}
\begin{tabularx}{\textwidth}{p{0.18\textwidth} X p{0.23\textwidth}}
\toprule
Institution & Role of expressed preferences & Reference \\
\midrule

French-speaking Belgium
&
Standard priority categories apply only at the student's first-choice school, and the composite tie-breaking index receives a larger weight when the school is ranked more highly. Then DA is used.
&
\citet{cantillon2015belgium} 
\\[4pt] \hline 

Taiwan high-school assignment
&
A student's admission score at a school is reduced according to the position at which the school is ranked.
Then DA is used.
&
\citet{dur2022deduction}
\\[4pt] \hline 

Germany, centralized university admissions
&
In the university-selection stage, universities could restrict consideration to applicants who ranked them sufficiently highly. For example, a university could consider only applicants who ranked it first and then rank those applicants according to academic criteria. (1)
&
\citet{westkamp2013}
\\[4pt] \hline

England
&
Applicants who ranked a school first received priority over applicants who ranked the same school lower.
These First Preference First arrangements were subsequently prohibited.
&
\citet{pathaksonmez2013}
\\[4pt] \hline

Trinity College, University of Toronto
&
Trinity considers for admission only students who select Trinity College as their first college preference.
&
\citet{trinitycollege}
\\[4pt] \hline

New York City high-school admissions
&
Under the historical Limited Unscreened admission method, applicants were selected randomly, but students who demonstrated interest in the program received higher priority.
&
\citet{nycdoe2016}
\\

\bottomrule
\end{tabularx}

\begin{minipage}{0.96\textwidth}
\footnotesize
\textit{Notes:}
(1) {Rank-based preselection was used by a substantial number of German universities in the 2010--11 admission cycle. For medicine, \citet{westkamp2013} reports that, out of 34 universities, 11 considered only applicants who ranked them first, while 5 and 5 imposed second- and third-rank thresholds, respectively. Similar restrictions were used in dentistry, pharmacy, and veterinary medicine; for example, 3 of the 5 veterinary-medicine programs considered only applicants who ranked them first.
See Table~1 in \citet{westkamp2013}.}
\end{minipage}
\end{table}

The desire to prioritize ``enthusiastic'' applicants is also evident in decentralized college admissions, where early-decision programs allow students to substantially improve their probability of admission \citep{avery2003early, avery2010early}. Finally, the recent work of \cite{rothstein2026school} argues for first-rank priorities as a tool to address school segregation.

In our model, a unit mass of students is to be assigned to a finite set of capacity-constrained schools (\cite{azevedo2016supply}). A student's type is  their preference ordering over schools, together with a von Neumann–Morgenstern utility representation of that ordering. Note that we assume a fixed and known ``cardinalization'' of each preference. Students' preferences are their private information, but for each preference there is a known utility representation. The planner's objective stems from a payoff for each assignment of a student to a school. We make no assumption on the relation between the planner's and the students' utilities. Thus, the planner may care about the rank of the assigned school, student (utilitarian) welfare, match quality, diversity, or any other characteristic that can be represented by type and school-dependent payoffs. We consider optimal direct-revelation mechanisms. A mechanism specifies, for every reported preference type and every school, the mass of students assigned to that school. The planner chooses an optimal assignment subject to school capacities and incentive compatibility.

Deferred acceptance provides the natural starting point for the analysis. We study both multiple tie-breaking (MTB), under which each student draws an independent lottery number at each school, and single tie-breaking (STB), under which the same lottery number determines priority at every school. \textit{Our main result provides a characterization of when the standard DA-MTB assignment solves the planner's problem. Under interior school cutoffs, DA-MTB is optimal if and only if, for each school, the planner's payoff from assigning a student to that school is independent of the student's preference type.} 

Therefore, DA-MTB will typically \textit{not} solve the planner's problem. The DA-MTB can be optimal when the planner values some schools uniformly more than others, but it ceases to be optimal as soon as the identity of the student assigned matters through their preferences. In particular, DA-MTB is not optimal when the planner values assigning students to schools that they rank highly. For example, if the planner wants to maximize the number of first-rank matches, the DA is suboptimal.

The key ingredient behind our result is the slack in the incentive-compatibility constraints of DA-MTB. With interior admissions cutoffs, every preference type strictly prefers the lottery induced by truthful reporting to the lottery induced by any alternative report.  If the planner values assigning a given school to one preference type more than to another, a local reassignment across those types can improve the planner's objective while preserving feasibility and incentive compatibility. Conversely, when the planner's payoff from each school is independent of preference type, every assignment that fills all school capacities attains the same maximal school-based payoff, and DA-MTB is optimal.

Given that the DA does not generally solve the planner's problem, we study how the planner can improve upon the DA by adjusting its admission cutoffs. Here we make progress under some restrictive, but hopefully illustrative, special cases. When there are two schools, we obtain a complete characterization within natural classes of rank-dependent cutoff adjustments. We suppose that the planner places greater weight on first-choice assignments. Starting from DA-MTB, we allow the cutoff faced by an applicant at a school to depend on whether they rank that school first or second. We consider adjustments that lower the cutoff faced by first-choice applicants and raise the cutoff faced by second-choice applicants while preserving school capacities. We characterize the optimal adjustments within this class. An analogous exercise starting from DA-STB is also performed. Moreover, we show that the optimal MTB and STB adjustments assign exactly the same total mass of students to their first choices (a result that stands in contrast to some findings in the literature that point to an advantage for STB). 

Finally, we turn to a model with many schools. We first connect abstract assignments with what can be achieved with scores and rank-dependent cutoffs. We show that any feasible assignment admits a cutoff representation under MTB once school priorities are allowed to depend on students' submitted preferences. However, this does not by itself imply implementation through student-proposing DA-MTB. When schools are allowed to deem some students unacceptable, we show that student-proposing DA then reproduces exactly the lottery associated with every possible report.

Second, we note that an optimal solution often has each type of student assigned to a small number of possible schools. We show that this sparsity, or small support, property holds generally for an approximate solution to the planner's problem.  We define an \(\epsilon\)-approximate solution as a feasible assignment whose objective value is within \(\epsilon\) of the planner's optimum and in which every incentive constraint can be violated by at most \(\epsilon\). Under equal type masses and an integrality condition on capacities, our approximation theorem shows that for every \(\epsilon>0\), there exists an \(\epsilon\)-approximate solution in which each represented preference type is assigned with positive probability to at most $1+\frac{4\ln|\mathrm{Types}|}{\epsilon^2}$ different schools; where $\mathrm{Types}$ is the set of possible student types. Moreover, for every represented type and every \(k\), the probability of receiving one of their top-\(k\) schools differs from that under the optimum by at most \(\epsilon\). In particular, when the number of types grows polynomially with the number of schools \(n\), each type needs to be assigned to only \(O(\ln n/\epsilon^2)\) schools. 

The remainder of the paper is organized as follows. Section~\ref{sec:literature} reviews the related literature. Section~\ref{model} presents the model and formulates the planner's problem. Section~\ref{sec:da} introduces deferred acceptance under single and multiple tie-breaking. Section~\ref{sec:da-optimal} presents our main characterization of when DA-MTB solves the planner's problem and establishes the strict incentive-compatibility property underlying that result. Section~\ref{sec:two-schools} characterizes optimal rank-dependent cutoff adjustments in the two-school environment. Section~\ref{sec:implementation} studies the representation and implementation of feasible assignments through endogenous school priorities. Section~\ref{Approx} establishes the sparse approximation result for many-school environments. Section~\ref{sec:conclusion} concludes.

\section{Literature Review}\label{sec:literature}

Our paper is related to several strands of the literature on school choice and matching. We first connect to the literature on the design of school-choice mechanisms. The deferred-acceptance (DA) algorithm was introduced by \citet{gale1962college} in the college-admissions problem and adapted to the school-choice setting by \citet{abdulkadiroglu2003school}. Student-proposing DA has subsequently become a central benchmark in both the theory and practice of school choice. In our two-school analysis, we take DA as the benchmark and ask whether its priority structure can be modified so that students receive more favorable treatment at schools that they rank more highly while preserving incentive compatibility. 

Particularly related are mechanisms in which a student's submitted preferences affect their priority. \citet{dur2022deduction} study the Taiwan assignment mechanism, in which points are deducted from an applicant's priority score according to the position of a school in their submitted preferences, and DA is then applied using the resulting priorities. They show that every nonzero deduction rule is manipulable and that, under their assumptions, the scope for manipulation increases with the size of the deductions. \citet{ayoadepapai2023} study Preference Rank Partitioned rules, a broad family of DA-based mechanisms in which schools' choices depend on both exogenous priority ranks and the positions at which students rank schools. They show that DA is the only strategy-proof rule within the class of rank-partition stable rules. \citet{sasaki2025} studies rank-dependent modifications of exogenous priorities and shows that stability with respect to the modified priorities and strategy-proofness are incompatible. Most recently, \citet{eguchi2026} allows a central planner to choose both the priority modification and the matching rule and shows that rank monotonicity, stability with respect to the modified priorities, and strategy-proofness jointly imply outcome equivalence to student-proposing DA under the original priorities.

Our approach differs from these papers in both the design problem and the restrictions imposed on the planner. \citet{dur2022deduction} take a rank-dependent priority adjustment as given and study its incentive consequences. \citet{ayoadepapai2023} retain exogenous priorities and work within the class of rank-partition stable rules, while \citet{sasaki2025} and \citet{eguchi2026} likewise take exogenous priorities as primitive and impose stability with respect to modified priorities together with restrictions on how submitted ranks may modify those priorities. We instead choose a feasible assignment directly subject to incentive compatibility and then ask how the resulting assignment can be represented and implemented through endogenous school priorities. Thus, we neither impose stability with respect to modified priorities nor require the report-dependent priorities used for implementation to satisfy rank monotonicity; indeed, the bonuses in our implementation need not be monotone in the reported rank of a school. These differences in the design problem and the implementation restrictions leave room for preference-dependent priorities to affect assignments while satisfying the incentive-compatibility constraints in our design problem.

\citet{abdulkadiroglu2015expanding} propose the choice-augmented deferred-acceptance mechanism (CADA). In CADA, students submit their ordinal ranking together with an auxiliary message identifying a target school, and targeting a school improves the student's position when ties at that school are broken. CADA preserves strategy-proofness with respect to the submitted ordinal ranking, but the targeting decision itself is strategic. Our approach differs in that we optimize directly over report-contingent assignment lotteries subject to feasibility and incentive compatibility, and only subsequently ask how the resulting assignments can be represented and implemented through report-dependent school priorities.

Our formulation is also related to \citet{abdulkadiroglu2021match}, who study the problem of optimizing policy objectives over stable school assignments under exogenously given priorities. Their main application maximizes aggregate match quality. Our problem differs in the constraint imposed on the planner and in the role of school priorities. We choose a feasible assignment subject to incentive compatibility and subsequently ask how the resulting assignment can be represented and implemented through report-dependent school priorities. 

A second connection is with the literature on endogenous and reciprocal preferences in matching markets. \citet{antler2015twosided} studies environments in which agents become more attractive when they express greater interest in potential partners, an idea related to our motivation for allowing school priorities to depend on submitted preferences. Antler also shows that acceptability may be essential when preferences are endogenous; similarly, we show that modifying relative school priorities alone may be insufficient for DA implementation, whereas allowing schools to place their outside option among students restores implementation. See also \citet{antler2019second} and \citet{opitz2023everyone} for related evidence and applications.

A third strand of related work compares the use of single and multiple tie-breaking under DA in terms of students' top choices. Empirical analyses of New York City and Amsterdam by \citet{abdulkadiroglu2009strategy} and \citet{de2023performance}, together with the theoretical analysis of \citet{ashlagi2019assigning}, show that STB often performs better in terms of top-choice assignments, while MTB may reduce lower-ranked assignments and the probability of remaining unmatched. Subsequent work emphasizes that this comparison depends on demand conditions (see, for example, \citep{ashlagi2020what} and \citep{allman2023rank}). Our comparison differs from this literature because we do not hold the priority structure fixed. Instead, we optimize rank-dependent cutoff adjustments separately around DA-STB and DA-MTB. Under the conditions of Proposition~\ref{prop:comp}, the two optimized adjustment classes attain the same total mass of first-choice assignments. 

Finally, our approximation result is methodologically related to applications of the probabilistic method in economics and algorithmic game theory. \citet{barman2015approximating} proves an approximate version of Carath\'eodory's theorem by expressing a point in a convex hull as a probability distribution over the generating vectors and using a concentration inequality. 
We use a related sampling idea to prove Theorem~\ref{thm:approx}. 

\section{The Model}\label{model}

A unit mass of atomistic students is to be matched to a finite set of $n$ schools. The set of students is the interval $[0,1]$. Let 
\(S=\{1,2,\dots,n\}\) denote the set of schools. Let
\[
q=(q_1,q_2,\dots,q_n)\in \mathbb{R}_+^n
\]
denote the vector of school capacities, where \(q_j>0\) for every \(j\in S\).

\paragraph{Students.}
Each student has a strict preference over schools, which we identify with the student's type. Let $\mathcal P$ denote the set of all strict preference orderings over $S\cup \{ \emptyset\}$, where the symbol \(\emptyset\) denotes the outside option of remaining unmatched. Suppose for simplicity that all schools are ranked above $\emptyset$. Then, $|\mathcal P|=n!$. Let $\eta\in \Delta(\mathcal P)$ be the distribution of student types. For each \(\succ\in \mathcal P\), \(\eta_\succ\) denotes the mass of students with preference \(\succ\). Let $\mcR := \{\succ \in \mathcal P \colon \eta_\succ>0\}$ be the support of \(\eta\). \footnote{We take $\mcR$ as the set of admissible reports for each student. Note that this is without loss of generality for our results if off-support reports are assigned the outside option with probability one.} 

Each ordinal preference \(\succ\in \mcR \) is associated with a fixed von Neumann–Morgenstern utility function $u_\succ:S\cup\{\emptyset\}\to[0,1]$  representing $\succ$, so  $j\succ j'$ if and only if $u_\succ(j)>u_\succ(j')$. Suppose that $u_\succ(\emptyset)=0$ for every $\succ\in\mcR$, a normalization. So $u_{\succ}(j)>0$ for every $(\succ,j)\in \mcR \times S$.

An important assumption in our model is that the cardinal utility representation for each ordinal preference is fixed and known. Preferences are students' private information, so we identify a student's type with their ordinal preference. We use the von Neumann–Morgenstern utility to evaluate lotteries over schools and note that eliciting cardinal utilities in environments without money and quasilinearity is very challenging. Hence we focus on settings with ordinal mechanisms and assume a known representation for each preference: for example, a utility of $1$ for the top ranked school, $(n-1)/n$ for the second-ranked, etc. 

\paragraph{Planner payoffs.}
For each \(\succ\in \mathcal P\), let $v_{\succ}:S\cup\{\emptyset\}\to [0,1]$ denote the planner's payoffs over a student of type $\succ$'s assignment: so \(v_{\succ}(j)\) is the planner's payoff when a student of type \(\succ\) is assigned to school \(j\). Suppose that $v_{\succ}(\emptyset)=0$. Thus, the planner weakly prefers assigning a student to a school over leaving them unmatched. At this point, we assume nothing about how \(u_\succ\) and \(v_\succ\) are related; the planner's objective need not coincide with student welfare. 

For example, $v_\succ(s)$ could equal $1$ if $s$ is the top-ranked school in $\succ$ and 0 otherwise. Such a payoff would capture a planner who wants to maximize the number of rank-1 assignments. Another example could be $v_\succ(s)=1$ for all $s$ and $v_\succ(\emptyset)=0$, which would model a planner who wants to match as many students as possible to a school. 

\paragraph{Assignments.}
A feasible assignment is a nonnegative matrix
\[ x = (x_{\succ,j})_{\succ\in \mcR,\;j\in S\cup\{\emptyset\}} \in \mathbb{R}_+^{|\mcR|\times(n+1)}
\]
such that, for all $\succ\in \mcR$ and all $j\in S$,
\begin{align*}
    \sum_{i\in S\cup\{\emptyset\}} x_{\succ,i}& =\eta_\succ, \\
    \sum_{\succ'\in \mcR} x_{\succ',j}& \le q_j.
\end{align*}

Here,  \(x_{\succ,j}\) is the mass of students with preference \(\succ\) who are assigned to school \(j\), and \(x_{\succ,\emptyset}\)  is the mass of students with preference \(\succ\) that remain unmatched. Let $X$ be the set of feasible assignments. 

For each \(\succ\in\mcR\), an assignment $x$ defines a probability distribution over schools and the option of remaining unmatched. Let 
\[
p_{\succ}(x):=
(p_{\succ,1},\dots, p_{\succ,n},p_{\succ,\emptyset}) := \left(\frac{x_{\succ,1}}{\eta_\succ},\dots,\frac{x_{\succ,n}}{\eta_\succ},\frac{x_{\succ,\emptyset}}{\eta_\succ}\right)\in \Delta(S\cup\{\emptyset\}).
\]

For every $\succ\in \mcR$, we extend $u_\succ$ and $v_\succ$ linearly to the lotteries induced by assignments in $X$. Given a feasible assignment $x\in X$, denote by $u_{\succ}(x)$ the expected utility of a student of type \(\succ\), and by \(v_\succ(x)\) the planner's expected payoff per student of type \(\succ\), under \(x\):
\[
u_{\succ}(x) = \sum_{j\in S} u_\succ(j)p_{\succ,j} \qquad v_\succ(x) = \sum_{j\in S}v_\succ(j)p_{\succ,j}.
\]

\paragraph{The planner's optimization problem.} The planner chooses a feasible assignment \(x\in X\) to maximize expected payoff $\sum_{\succ\in\mcR}\sum_{j\in S} v_\succ(j)x_{\succ,j}$. A student of type $\succ$ should be better off (or no worse off) with their intended assignment in $x$ than with the assignment they would receive by reporting a different type $\succ'$. So we want that 
$\sum_{j\in S}u_\succ(j)p_{\succ,j} \geq \sum_{j\in S}u_\succ(j)p_{\succ',j}$; where $p_{\succ,j}=\frac{x_{\succ,j}}{\eta_\succ}$. Thus, the planner solves
\begin{equation}\label{problem}
\begin{aligned}
\max_{x\in X}\quad
& \sum_{\succ\in\mcR}\sum_{j\in S} v_\succ(j)x_{\succ,j} \\[3pt]
\text{s.t.}\quad
& \sum_{j\in S}u_\succ(j)\frac{x_{\succ,j}}{\eta_\succ} \geq \sum_{j\in S}u_\succ(j)\frac{x_{\succ',j}}{\eta_{\succ'}},
\qquad
\text{for all }\succ,\succ'\in\mcR.
\end{aligned}
\end{equation}

Problem~\eqref{problem} is a linear program. Its feasible set is nonempty (for example, assigning every type to the outside option satisfies feasibility and incentive compatibility) and compact. Hence, an optimal solution exists.

\section{The Deferred Acceptance Mechanism}\label{sec:da}

We first describe the student-proposing Deferred Acceptance outcome under random priorities, following \cite{azevedo2016supply}. We consider two priority structures. Under deferred acceptance with single tie-breaking (DA-STB), each student draws one lottery number that determines their priority at every school. Under deferred acceptance with multiple tie-breaking (DA-MTB), each student draws an independent lottery number for each school. In both cases, a higher lottery number indicates higher priority. A school's cutoff is the lowest lottery number that makes a student eligible for admission to that school.

\paragraph{The deferred acceptance algorithm.}
Fix a realization of priorities and a profile of reported ordinal preferences. Under the student-proposing deferred acceptance algorithm, each student first applies to their most preferred school according to their reported preferences. Each school tentatively admits, among its applicants, the students with highest priority up to its capacity, and rejects the rest. At every subsequent step, each rejected student applies to their most preferred school according to their reported preferences among those that have not yet rejected them. Each school then considers its new applicants together with the students tentatively admitted in the previous step, retains the highest-priority students up to capacity, and rejects the rest. The process continues until no new applications are made. At that point, all tentative assignments become final, and any student who has been rejected by every school remains unmatched.

For each reported type \(\succ\in\mcR\), let
\(\pi_k(\succ)\) denote the school ranked in position
\(k\in\{1,\ldots,n\}\) under \(\succ\).

\subsection{DA-STB}

Under deferred acceptance with single tie-breaking, each student independently draws a single lottery number $\omega\sim U[0,1]$. The same lottery number determines their priority at every school. The DA-STB outcome is characterized by a vector of school cutoffs \(\{c^{STB}_s\}_{s\in S}\), where \(c^{STB}_s\in[0,1]\) is the lowest lottery number required for admission at school \(s\) \citep{azevedo2016supply}. Thus, a student is eligible for admission to school $s$ if and only if $\omega\geq c^{STB}_s$. They are assigned to their most preferred school among those for which they are eligible. Therefore, the DA-STB allocation can be represented as a market-clearing cutoff allocation in the sense of \citet{azevedo2016supply}.

Under DA-STB, a student reporting \(\succ\) is assigned to school \(\pi_k(\succ)\) if and only if their lottery number makes them eligible for admission there, but does not make them eligible for admission to any school that they prefer more. That is,
\[
c^{STB}_{\pi_k(\succ)}\le \omega < \min_{\ell<k} c^{STB}_{\pi_\ell(\succ)},
\]
with the convention that \(\min\varnothing=1\). If we consider the mechanism ex-ante, before the realization of the lottery numbers, we have an assignment as defined in the previous section. Let $p^{STB}_\succ$ denote the ex ante lottery induced by DA-STB for reported type \(\succ\). Then,
\[
p^{STB}_{\succ,\pi_k(\succ)} = \max\left\{ \min_{\ell<k}c^{STB}_{\pi_\ell(\succ)} - c^{STB}_{\pi_k(\succ)}, \,0 \right\}.
\]
A student reporting \(\succ\) remains unmatched if their lottery number is below every school cutoff, so $p^{STB}_{\succ,\emptyset} = \min_{j\in S} c^{STB}_j$. Let $x^{STB}$ denote the assignment induced by DA-STB. Then $x^{STB}_{\succ,j} = \eta_\succ p^{STB}_{\succ,j}$, for all $(\succ,j)\in\mcR\times (S\cup \{\emptyset\})$.

\subsection{DA-MTB}
Under deferred acceptance with multiple tie-breaking, each student draws a lottery number $\omega_s\in U[0,1]$ that determines their priority at school $s$. The draws are independent across schools and students. The DA-MTB outcome is characterized by a vector of school cutoffs \(\{ c^{MTB}_s\}_{s\in S}\) \citep{azevedo2016supply}. 

Under DA-MTB, a student reporting \(\succ\) is assigned to school
\(\pi_k(\succ)\) if and only if they are ineligible for admission to every school they prefer to \(\pi_k(\succ)\), but eligible for admission to \(\pi_k(\succ)\). Since lottery numbers are independent across schools, this occurs if and only if
\[
\omega_{\pi_\ell(\succ)}< c^{MTB}_{\pi_\ell(\succ)}
\quad\text{for every }\ell<k,
\qquad\text{and}\qquad
\omega_{\pi_k(\succ)}\ge  c^{MTB}_{\pi_k(\succ)}.
\]

Let $p^{MTB}_\succ$ denote the ex ante lottery induced by DA-MTB for a reported type \(\succ\). So
\[
p^{MTB}_{\succ,\pi_k(\succ)}
=
\bigl(1- c^{MTB}_{\pi_k(\succ)}\bigr)\prod_{\ell<k} c^{MTB}_{\pi_\ell(\succ)}.
\]
A student reporting \(\succ\) remains unmatched if they are rejected by every school, so
\[
p^{MTB}_{\succ,\emptyset} = \prod_{\ell=1}^n  c^{MTB}_{\pi_\ell(\succ)}=\prod_{s\in S}c_s^{MTB}.
\]

Let $x^{MTB} = \left( x^{MTB}_{\succ,j} \right)_{\succ\in\mcR,\; j\in S\cup\{\emptyset\}}$ denote the assignment induced by DA-MTB. Then $x^{MTB}_{\succ,j} = \eta_\succ p^{MTB}_{\succ,j}$ for all $(\succ,j)\in \mcR\times (S\cup\{\emptyset\})$. 

Since the cutoffs are market-clearing, the assignments satisfy
\[
\sum_{\succ\in\mcR}\eta_\succ p^{\tau}_{\succ,j}
\le q_j
\qquad
\text{for every }j\in S,
\]
for each \(\tau\in\{STB,MTB\}\). Hence, the corresponding assignments \(x^{\tau}\) are feasible.

\subsection{Properties of DA.} In our continuum economy, individual students have measure zero. The finite-economy notion of a blocking pair is replaced by a cutoff-based stability condition. We distinguish between the ex ante assignments defined in Section~\ref{model} and realized assignments. An ex ante assignment \(x\in X\) specifies type-by-school masses before the realization of the lottery numbers. By contrast, after priorities are realized, a realized assignment is a measurable function $\mu:[0,1]\to S\cup\{\emptyset\}$, where \(\mu(i)\) denotes the school assigned to student \(i\), or \(\emptyset\) if they remain unmatched. A realized assignment is feasible if, for each school, the measure of students assigned to that school does not exceed its capacity.

Fix a realization of priorities. A feasible realized assignment \(\mu\) is \textit{stable} if there exists a vector of cutoffs
\[
c=(c_s)_{s\in S}, \qquad c_s\in[0,1],
\]
such that the following conditions hold. First, for almost every student \(i\), \(\mu(i)\) is the most preferred school among those schools \(s\) for which student \(i\) is eligible; if no such school exists, student \(i\) remains unmatched. Second, for every school \(s\), the mass of students assigned to \(s\) is at most \(q_s\). Third, if school \(s\) has unfilled capacity, then \(c_s=0\); equivalently, a positive cutoff can occur only at a school whose capacity is filled. Under this definition, deferred acceptance yields the student-optimal stable cutoff allocation; in finite economies, this coincides with the classical student-optimal stable matching \citep{gale1962college, roth1982economics}.

Both DA-STB and DA-MTB depend only on students' ordinal preferences and school priorities. Truthful reporting of ordinal preferences is a dominant strategy in the finite model, and both mechanisms remain ordinally strategy-proof in the large economy \citep{dubins1981machiavelli, roth1982economics, abdulkadiroglu2015expanding}. That is, for every true preference \(\succ\), the lottery induced by reporting \(\succ\) first-order stochastically dominates, with respect to \(\succ\), the lottery induced by any other report.

\section{When does DA-MTB solve the planner's problem?}
\label{sec:da-optimal}

We first ask when the DA-MTB assignment solves the Problem~\eqref{problem}. Our first main result provides a characterization under interior cutoffs: DA-MTB is optimal if and only if, for each school, the planner's payoff from assigning a student to that school is independent of the student's preference type. This is essentially a negative result. The planner may care about which school a student is assigned to, but not about where that school appears in the student's preference ranking. In particular, for the DA-MTB to be optimal, assigning a student to a given school must give the same payoff to the planner, whether that school is the student's first choice, last choice, or any intermediate choice.

\begin{theorem}\label{thm:da-optimal}
Suppose that all DA-MTB cutoffs are interior: $c_j^{MTB}\in(0,1)$ for every $j\in S$.
Then the DA-MTB assignment is an optimal solution to Problem~\ref{problem} if and only if, for every school \(j\in S\),
\[
v_\succ(j)=v_{\succ'}(j)
\qquad
\text{for every }\succ,\succ'\in\mcR.
\]
\end{theorem}

Theorem~\ref{thm:da-optimal} (whose proof is in Appendix~\ref{proof:cor:da}) states that the DA-MTB is suboptimal as soon as the planner values the placement of a student in a school depending on the student's preferences. If, for some school, assigning that school to one preference type yields a different payoff than assigning it to another type, DA-MTB is no longer optimal. A common situation in practice is a rank-based objective, under which the planner values assignments more highly when students receive schools that they rank more highly. In such a situation, DA-MTB is suboptimal. This observation motivates our analysis of assignments that improve upon DA while preserving incentive compatibility.

The key ingredient behind Theorem~\ref{thm:da-optimal} is the following strict incentive-compatibility property of DA-MTB. Under interior cutoffs, every type strictly prefers the lottery induced by truthful reporting to the lottery induced by any other report. Consequently, the DA-MTB assignment has local slack in every incentive-compatibility constraint.

\begin{lemma}\label{lem:da-strict-ic}
Suppose that all DA-MTB cutoffs are interior: $c_j^{MTB}\in(0,1)$ for every $j\in S$.
Then, for every true type \(\succ\in\mcR\) and every alternative report
\(\succ'\in\mcR\setminus\{\succ\}\),
\[
\sum_{j\in S}
u_\succ(j)p^{MTB}_{\succ,j}
>
\sum_{j\in S}
u_\succ(j)p^{MTB}_{\succ',j}.
\]
\end{lemma}

The proof of Lemma~\ref{lem:da-strict-ic} is in Appendix~\ref{DA:proof}. The interior-cutoff assumption is important for Lemma~\ref{lem:da-strict-ic}.\footnote{If some DA-MTB cutoff lies on the boundary, strict incentive compatibility may fail. For example, suppose there are three schools \(a,b,c\) and \(c_a^{MTB}=0\). Since \(a\) is then always attainable, the reports \(a\succ b\succ c\) and \(a\succ c\succ b\) induce the same assignment whenever both rank \(a\) first. Thus the corresponding incentive constraint binds.}

\begin{remark}
The strictness conclusion in Lemma~\ref{lem:da-strict-ic} need not hold under STB. Under STB, different reports may induce exactly the same lottery. For example, under DA-STB with three schools \(a,b,c\), suppose $0<c_a^{STB}<c_b^{STB}<1$, and $0<c_a^{STB}<c_c^{STB}<1$. Consider two reports that both rank \(a\) first but differ only in the ordering of \(b\) and \(c\). A student using either report is assigned to \(a\) whenever \(\omega\geq c_a^{STB}\), and is unmatched whenever \(\omega<c_a^{STB}\), because in the latter case $\omega<c_a^{STB}<c_b^{STB},c_c^{STB}$. Thus the two reports induce the same lottery, so the corresponding incentive constraint binds.
\end{remark}

\section{The case of two schools}
\label{sec:two-schools}

We shall first discuss a simple model of two schools under the assumption that the planner cares only about the rank of the assignment. We do not solve the unrestricted version of Problem~\eqref{problem}; rather, we characterize optimal adjustments within the MTB and STB adjustment classes defined below. Denote the two schools by  \(A\) and \(B\), with capacities \(q_A,q_B>0\). Let \(\succ_A\) denote the preference ordering \(A\succ B\), and \(\succ_B\) be \(B\succ A\). A mass $\eta_A:=\eta_{\succ_A}\in(0,1)$ of students has type \(\succ_A\), while  $\eta_B:=\eta_{\succ_B}=1-\eta_A$ has type \(\succ_B\). 

We assume that students obtain utility $u_1$ from their first choice and $u_2$ from their second choice with $1\geq u_1>u_2>0$. The planner obtains a payoff \(\alpha\in(1/2,1]\) from a first-choice assignment, \(1-\alpha\) from a second-choice assignment, and zero from leaving a student unmatched. As shown below, within the full-capacity adjustment classes we consider, maximizing the planner's payoff is equivalent to maximizing the mass of first-choice assignments.

\subsection{DA-MTB}
\begin{figure}[h!]
    \centering
    \tikzset{every picture/.style={line width=0.75pt}} 

\begin{tikzpicture}[x=0.75pt,y=0.75pt,yscale=-1,xscale=1,scale=0.7]

\draw   (33,72) -- (285.2,72) -- (285.2,291.6) -- (33,291.6) -- cycle ;
\draw    (179,73) -- (180.2,295.6) ;
\draw    (33,221) -- (287.2,220.6) ;
\draw   (346,73) -- (598.2,73) -- (598.2,292.6) -- (346,292.6) -- cycle ;
\draw    (492,74) -- (493.2,296.6) ;
\draw    (346,222) -- (600.2,221.6) ;

\draw (126,43.4) node [anchor=north west][inner sep=0.75pt]    {$A\succ B$};
\draw (171,297.4) node [anchor=north west][inner sep=0.75pt]    {$c_{A}$};
\draw (10,209.4) node [anchor=north west][inner sep=0.75pt]    {$c_{B}$};
\draw (24,296.4) node [anchor=north west][inner sep=0.75pt]    {$0$};
\draw (280,295.4) node [anchor=north west][inner sep=0.75pt]    {$1$};
\draw (16,62.4) node [anchor=north west][inner sep=0.75pt]    {$1$};
\draw (215,138.4) node [anchor=north west][inner sep=0.75pt]    {$A$};
\draw (216,245.4) node [anchor=north west][inner sep=0.75pt]    {$A$};
\draw (91,134.4) node [anchor=north west][inner sep=0.75pt]    {$B$};
\draw (102,245.4) node [anchor=north west][inner sep=0.75pt]    {$\emptyset $};
\draw (439,44.4) node [anchor=north west][inner sep=0.75pt]    {$B\succ A$};
\draw (484,298.4) node [anchor=north west][inner sep=0.75pt]    {$c_{A}$};
\draw (323,210.4) node [anchor=north west][inner sep=0.75pt]    {$c_{B}$};
\draw (337,297.4) node [anchor=north west][inner sep=0.75pt]    {$0$};
\draw (593,296.4) node [anchor=north west][inner sep=0.75pt]    {$1$};
\draw (329,63.4) node [anchor=north west][inner sep=0.75pt]    {$1$};
\draw (528,139.4) node [anchor=north west][inner sep=0.75pt]    {$B$};
\draw (529,246.4) node [anchor=north west][inner sep=0.75pt]    {$A$};
\draw (404,135.4) node [anchor=north west][inner sep=0.75pt]    {$B$};
\draw (415,246.4) node [anchor=north west][inner sep=0.75pt]    {$\emptyset $};

\end{tikzpicture}
    \caption{\footnotesize{DA-MTB assignment for the two preference types. The unit square represents the independent lottery draws \((\omega_A,\omega_B)\), and the regions indicate whether a student is assigned to \(A\), to \(B\), or remains unmatched given the cutoffs \(c_A\) and \(c_B\).}}
    \label{fig9}
\end{figure}
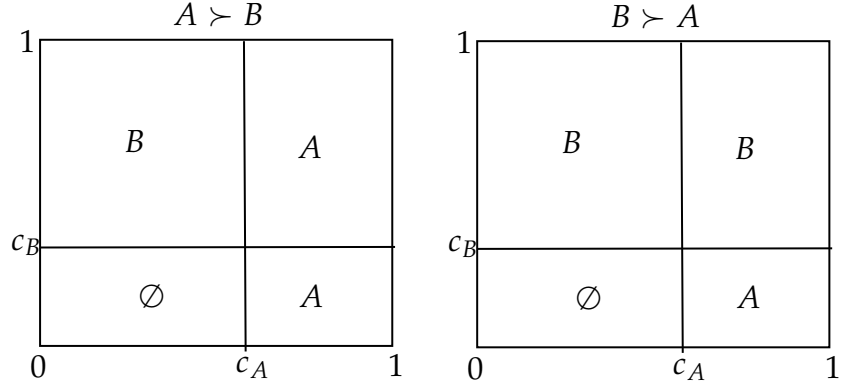

Under DA-MTB, each student draws two independent lottery numbers \((\omega_A,\omega_B)\in[0,1]^2\) from the uniform distribution, where \(\omega_j\) determines priority at school \(j\). School $j$ has a cutoff $c_j$ and admits all students with $\omega_j\geq c_j$, $j=A,B$. If both schools are filled at the DA-MTB outcome, the cutoffs are characterized by
\begin{align*}
\eta_A(1-c_A)+\eta_B c_B(1-c_A)& =q_A,   \\
\eta_B(1-c_B)+\eta_A c_A(1-c_B)& =q_B. 
\end{align*}
Figure \ref{fig9} illustrates the DA-MTB assignment.

\paragraph{Adjustments to DA-MTB.} The planner now chooses cutoffs that depend on the reported rankings. A rank-dependent cutoff mechanism in the two-school MTB environment is specified by cutoffs
\[
c_A^1,c_A^2,c_B^1,c_B^2\in[0,1],
\]
where \(c_j^1\) is the cutoff faced by students who rank school \(j\) first, and \(c_j^2\) is the cutoff faced by students who rank school \(j\) second.


Incentive compatibility requires that 
\begin{align}
(1-c_A^1)u_1+c_A^1(1-c_B^2)u_2 &\ge c_B^1(1-c_A^2)u_1+(1-c_B^1)u_2, \tag{$IC_A$} \\ 
(1-c_B^1)u_1+c_B^1(1-c_A^2)u_2 &\ge c_A^1(1-c_B^2)u_1+(1-c_A^1)u_2. \tag{$IC_B$} 
\end{align}
where $IC_A$ ensures that a student of type $A\succ B$ prefers to report this preference over the opposite preference, and $IC_B$ similarly ensures truth-telling for students of type $B\succ A$. The adjusted cutoffs are feasible if the induced assignment does not exceed school capacities: $\eta_A(1-c_A^1)+\eta_B c_B^1(1-c_A^2)\le q_A$ and  $\eta_B(1-c_B^1)+\eta_A c_A^1(1-c_B^2)\le q_B$. For the remainder of this subsection, we assume $q_A+q_B\leq 1$. Since all schools are acceptable, both schools fill their capacities at the market-clearing DA-MTB outcome. We further restrict attention to rank-dependent adjustments that preserve full capacity at both schools.

To illustrate the deviation from the DA, we formulate the planner's problem as choosing adjustments to the DA cutoffs. We lower the cutoff for first-choice applicants by $\epsilon$ or $\delta$  and raise the cutoff for second-choice applicants by $\epsilon'$ or $\delta'$:
\[
c_A^1=c_A-\epsilon,\qquad c_A^2=c_A+\epsilon',\qquad c_B^1=c_B-\delta,\qquad c_B^2=c_B+\delta',
\]
where we require that $0\leq\epsilon\leq c_A$, $0\leq\delta\leq c_B$, $0\leq\epsilon'\leq1-c_A$, and  $0\leq\delta'\leq1-c_B$.

We further restrict attention to adjustments that preserve full capacity at both schools. 
Within this class, the total mass of assigned students is fixed
at \(q_A+q_B\). Hence, if \(F\) denotes the mass of students assigned to their first choice, the planner's payoff can be written as
\[
\alpha F+(1-\alpha)(q_A+q_B-F)=(1-\alpha)(q_A+q_B)+(2\alpha-1)F.
\]
Because \(\alpha>1/2\), maximizing the planner's payoff is equivalent to maximizing \(F\). The planner's problem over this class of MTB adjustments is:
\begin{equation}
    \begin{aligned}
\max_{\epsilon,\delta,\epsilon',\delta'} 
\quad 
& \eta_A(1-c_A+\epsilon) + \eta_B(1-c_B+\delta) \\[0.3em]
\text{s.t.} \quad 
& \epsilon \eta_A 
= 
\eta_B\left[\epsilon'(c_B-\delta)+\delta(1-c_A)\right],
\qquad \\[0.3em]
& \delta \eta_B 
= 
\eta_A\left[\epsilon(1-c_B)+\delta'(c_A-\epsilon)\right],
\qquad  \\[0.3em]
& u_1(1-c_A+\epsilon)
+u_2(c_A-\epsilon)(1-c_B-\delta')  \\
&\hspace{3em}\geq 
u_1(1-c_A-\epsilon')(c_B-\delta)
+u_2(1-c_B+\delta), \\[0.3em]
& u_1(1-c_B+\delta)
+u_2(c_B-\delta)(1-c_A-\epsilon')  \\
&\hspace{3em}\geq 
u_1(c_A-\epsilon)(1-c_B-\delta')
+u_2(1-c_A+\epsilon), \\[0.3em]
& \epsilon \leq c_A, \\ 
& \delta' \leq 1-c_B, \\ 
& \epsilon' \leq 1-c_A, \\ 
& \delta \leq c_B, \\ 
& \epsilon,\epsilon',\delta,\delta' \geq 0.
\end{aligned}
\label{eq:planner-problem-2schools2}
\end{equation}

Since the feasible set is compact and the objective is continuous, a solution exists. The DA-MTB assignment corresponds to setting $\epsilon=\epsilon'=\delta=\delta'=0,$ which is feasible for the planner's problem.

Because lowering first-choice cutoffs increases the mass of first-choice assignments, the planner would like to increase both \(\epsilon\) and \(\delta\) as much as possible. Proposition~\ref{2schools} shows that an optimum is therefore determined either by the maximal feasible adjustment or by one of the two incentive constraints.

\begin{proposition}\label{2schools}
Let
\[ \bar\epsilon:= \min\left\{ c_A,\, \frac{\eta_B}{\eta_A}(1-c_A)c_B \right\}, \qquad \bar\delta:= \min\left\{ c_B,\, \frac{\eta_A}{\eta_B}(1-c_B)c_A \right\}.
\]
(These are the maximal values of \(\epsilon\) and \(\delta\) compatible with
the full-capacity conditions and the cutoff bounds.)

Then any optimal MTB adjustment $(\epsilon^*,\delta^*,\epsilon^{\prime *},\delta^{\prime *})$ must be of exactly one of the following three types:
\begin{enumerate}
    \item[\textup{(i)}] \(IC_A\) binds, \(IC_B\) is slack, and
    \[
    \epsilon^*=\bar\epsilon,
    \]
    \[
\delta^*=\frac{u_1}{u_2}\frac{\eta_A}{\eta_B}\bar\epsilon+(1-c_A)(1-c_B)\left(\frac{u_1}{u_2}-1\right)\eta_A.
\]
    \item[\textup{(ii)}] both incentive constraints are slack, $\epsilon^*=\bar\epsilon$, and $\delta^*=\bar\delta.$
    \item[\textup{(iii)}] \(IC_A\) is slack, \(IC_B\) binds, $\delta^*=\bar\delta$, and 
    \[
\epsilon^*=\left[\bar\delta+(1-c_A)(1-c_B)\left(1-\frac{u_2}{u_1}\right)\eta_A\right]\frac{u_1}{u_2}\frac{\eta_B}{\eta_A}.
\]
\end{enumerate}
The corresponding values of \(\epsilon^{\prime *}\) and \(\delta^{\prime *}\) are determined by the full-capacity conditions.
\end{proposition}

\textit{Proof sketch.}
Using the full-capacity conditions, the problem can be reduced to one in
\((\epsilon,\delta)\). The cutoff and capacity restrictions imply
\[
0\leq\epsilon\leq\bar\epsilon,
\qquad
0\leq\delta\leq\bar\delta,
\]
together with two additional linear restrictions relating \(\epsilon\) and \(\delta\). As shown in Appendix \ref{proof_2schools}, the northeast corner \((\bar\epsilon,\bar\delta)\) satisfies these additional restrictions. Since the objective is strictly increasing in both variables, the planner prefers northeast movements whenever they remain incentive compatible. The two IC boundaries cannot bind simultaneously, so either the northeast corner is the optimal MTB adjustment or the optimum lies on exactly one of the two incentive-compatibility boundaries, yielding the three cases stated in the proposition. A complete characterization together with the closed-form solution for \((\epsilon^*,\delta^*)\) is given in Appendix \ref{proof_2schools}.

Figure \ref{fig:cases123} illustrates the three possible cases.

\begin{figure}[ht]
\centering

\begin{subfigure}[b]{0.31\textwidth}
    \centering
    \resizebox{\linewidth}{!}{\tikzset{every picture/.style={line width=0.75pt}} 

\begin{tikzpicture}[x=0.75pt,y=0.75pt,yscale=-1,xscale=1, scale=0.8]

\draw    (177,252) -- (458,252) ;
\draw [shift={(460,252)}, rotate = 180] [color={rgb, 255:red, 0; green, 0; blue, 0 }  ][line width=0.75]    (10.93,-3.29) .. controls (6.95,-1.4) and (3.31,-0.3) .. (0,0) .. controls (3.31,0.3) and (6.95,1.4) .. (10.93,3.29)   ;
\draw    (185,259) -- (185,33) ;
\draw [shift={(185,31)}, rotate = 90] [color={rgb, 255:red, 0; green, 0; blue, 0 }  ][line width=0.75]    (10.93,-3.29) .. controls (6.95,-1.4) and (3.31,-0.3) .. (0,0) .. controls (3.31,0.3) and (6.95,1.4) .. (10.93,3.29)   ;
\draw    (178,187) -- (190,187) ;
\draw    (238,245) -- (238,261) ;
\draw    (186,187) -- (279,23) ;
\draw    (239,253) -- (393,137) ;
\draw    (226,52) -- (225,256) ;
\draw    (167,90) -- (316,90) ;
\draw    (186,228) -- (204,250) ;
\draw    (186,198) -- (224,248) ;
\draw    (194,178) -- (226,219) ;
\draw    (201,160) -- (226,186) ;
\draw    (210,147) -- (225,163) ;
\draw    (218,133) -- (225,142) ;

\draw (467,246.4) node [anchor=north west][inner sep=0.75pt]    {$\epsilon $};
\draw (157,22.4) node [anchor=north west][inner sep=0.75pt]    {$\delta $};
\draw (284,14.4) node [anchor=north west][inner sep=0.75pt]    {$IC_{A}$};
\draw (406,124.4) node [anchor=north west][inner sep=0.75pt]    {$IC_{B}$};
\draw (148,78.4) node [anchor=north west][inner sep=0.75pt]    {$\overline{\delta }$};
\draw (219,259.4) node [anchor=north west][inner sep=0.75pt]    {$\overline{\epsilon }$};

\end{tikzpicture}}
    \caption{Case I}
    \label{fig:case1}
\end{subfigure}
\hfill
\begin{subfigure}[b]{0.31\textwidth}
    \centering
    \resizebox{\linewidth}{!}{\tikzset{every picture/.style={line width=0.75pt}} 

\begin{tikzpicture}[x=0.75pt,y=0.75pt,yscale=-1,xscale=1,scale=0.8]

\draw    (177,252) -- (458,252) ;
\draw [shift={(460,252)}, rotate = 180] [color={rgb, 255:red, 0; green, 0; blue, 0 }  ][line width=0.75]    (10.93,-3.29) .. controls (6.95,-1.4) and (3.31,-0.3) .. (0,0) .. controls (3.31,0.3) and (6.95,1.4) .. (10.93,3.29)   ;
\draw    (185,259) -- (185,33) ;
\draw [shift={(185,31)}, rotate = 90] [color={rgb, 255:red, 0; green, 0; blue, 0 }  ][line width=0.75]    (10.93,-3.29) .. controls (6.95,-1.4) and (3.31,-0.3) .. (0,0) .. controls (3.31,0.3) and (6.95,1.4) .. (10.93,3.29)   ;
\draw    (178,187) -- (190,187) ;
\draw    (238,245) -- (238,261) ;
\draw    (186,187) -- (279,23) ;
\draw    (239,253) -- (393,137) ;
\draw    (326,46) -- (325,250) ;
\draw    (172,138) -- (336,138) ;
\draw    (188,224) -- (209,250) ;
\draw    (186,201) -- (224,251) ;
\draw    (194,178) -- (248,242) ;
\draw    (201,160) -- (262,232) ;
\draw    (210,147) -- (273,223) ;
\draw    (222,140) -- (284,217) ;
\draw    (190,187) -- (239,253) ;
\draw    (242,140) -- (296,210) ;
\draw    (262,140) -- (307,199) ;
\draw    (281,139) -- (318,192) ;
\draw    (299,139) -- (324,176) ;
\draw    (311,139) -- (325,160) ;

\draw (467,246.4) node [anchor=north west][inner sep=0.75pt]    {$\epsilon $};
\draw (157,22.4) node [anchor=north west][inner sep=0.75pt]    {$\delta $};
\draw (284,14.4) node [anchor=north west][inner sep=0.75pt]    {$IC_{A}$};
\draw (406,124.4) node [anchor=north west][inner sep=0.75pt]    {$IC_{B}$};
\draw (153,132.4) node [anchor=north west][inner sep=0.75pt]    {$\overline{\delta }$};
\draw (320.5,255.4) node [anchor=north west][inner sep=0.75pt]    {$\overline{\epsilon }$};

\end{tikzpicture}}
    \caption{Case II}
    \label{fig:case2}
\end{subfigure}
\hfill
\begin{subfigure}[b]{0.31\textwidth}
    \centering
    \resizebox{\linewidth}{!}{\tikzset{every picture/.style={line width=0.75pt}} 

\begin{tikzpicture}[x=0.75pt,y=0.75pt,yscale=-1,xscale=1, scale=0.8]

\draw    (177,252) -- (458,252) ;
\draw [shift={(460,252)}, rotate = 180] [color={rgb, 255:red, 0; green, 0; blue, 0 }  ][line width=0.75]    (10.93,-3.29) .. controls (6.95,-1.4) and (3.31,-0.3) .. (0,0) .. controls (3.31,0.3) and (6.95,1.4) .. (10.93,3.29)   ;
\draw    (185,259) -- (185,33) ;
\draw [shift={(185,31)}, rotate = 90] [color={rgb, 255:red, 0; green, 0; blue, 0 }  ][line width=0.75]    (10.93,-3.29) .. controls (6.95,-1.4) and (3.31,-0.3) .. (0,0) .. controls (3.31,0.3) and (6.95,1.4) .. (10.93,3.29)   ;
\draw    (178,187) -- (190,187) ;
\draw    (238,245) -- (238,261) ;
\draw    (186,187) -- (279,23) ;
\draw    (239,253) -- (393,137) ;
\draw    (338,60) -- (337,264) ;
\draw    (177,198) -- (341,198) ;
\draw    (188,224) -- (209,250) ;
\draw    (186,201) -- (224,251) ;
\draw    (211,198) -- (246.2,242.6) ;
\draw    (196,197) -- (234.2,248.6) ;
\draw    (229,199) -- (257.2,234.6) ;
\draw    (243,200) -- (268.2,231.6) ;
\draw    (259,198) -- (277.2,223.6) ;
\draw    (274,197) -- (287.2,216.6) ;
\draw    (289,199) -- (297.2,211.6) ;

\draw (467,246.4) node [anchor=north west][inner sep=0.75pt]    {$\epsilon $};
\draw (157,22.4) node [anchor=north west][inner sep=0.75pt]    {$\delta $};
\draw (284,14.4) node [anchor=north west][inner sep=0.75pt]    {$IC_{A}$};
\draw (406,124.4) node [anchor=north west][inner sep=0.75pt]    {$IC_{B}$};
\draw (160,185.4) node [anchor=north west][inner sep=0.75pt]    {$\overline{\delta }$};
\draw (329.5,265.4) node [anchor=north west][inner sep=0.75pt]    {$\overline{\epsilon }$};

\end{tikzpicture}}
    \caption{Case III}
    \label{fig:case3}
\end{subfigure}

\caption{\footnotesize{The three possible cases. Case I: \(IC_A\) binds, \(IC_B\) is slack, and \(\epsilon=\bar\epsilon\). Case II: both incentive constraints are slack, and \((\epsilon,\delta)=(\bar\epsilon,\bar\delta)\). Case III: \(IC_A\) is slack, \(IC_B\) binds, and \(\delta=\bar\delta\).}}
\label{fig:cases123}
\end{figure}
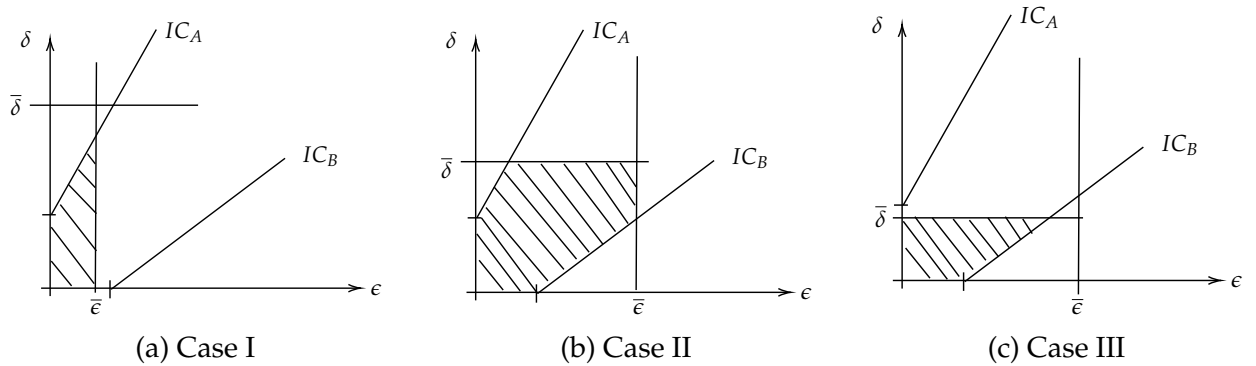


\subsection{DA-STB}
\label{sec:two-schools-stb}

We now consider adjustments to the DA-STB cutoffs. Under STB, the same lottery number determines priority at both schools. When school $A$ has greater first-choice demand per seat than school $B$, all seats at $A$ are assigned to students who rank it first, while some students who rank $A$ first are assigned to $B$. The planner can therefore improve the assignment by reallocating seats at $B$ to students who rank it first. We shall show that an optimum can be attained without changing the assignment at $A$, and characterize the optimal adjustment at $B$.

Let $c_A$ and $c_B$ be the DA-STB cutoffs. We focus on the case $q_A+q_B<1$ and $\eta_A/q_A>\eta_B/q_B$, so school $A$ is ``overdemanded.'' In consequence, $1>c_A>c_B>0$: insufficient aggregate capacity implies $c_B>0$, while $c_A>c_B$ if and only if $\eta_A/q_A>\eta_B/q_B$. 

School $A$ is filled entirely with students who rank it first, while $B$ admits both students who rank it first and students rejected from $A$. The capacity conditions are therefore $\eta_A(1-c_A)=q_A$ and $\eta_B(1-c_B)+\eta_A(c_A-c_B)=q_B$. Hence, $c_A=1-q_A/\eta_A$ and $c_B=1-q_A-q_B$. If $c_A=c_B$, no student is assigned to their second choice. If $c_B=0$, every student who ranks $B$ first is already admitted to $B$. Since we focus on interior cutoffs, there is scope for the reallocation of seats at $B$.

\begin{figure}[ht]
    \centering
    \tikzset{every picture/.style={line width=0.75pt}} 

\begin{tikzpicture}[x=0.75pt,y=0.75pt,yscale=-1,xscale=1]

\draw    (400,46) -- (402.29,304.49) ;
\draw    (251,49.51) -- (252.29,303) ;
\draw    (239.86,49.26) -- (262.14,49.75) ;
\draw    (388.86,45.75) -- (411.14,46.25) ;
\draw    (239.29,147.49) -- (264.29,147.49) ;
\draw   (230.29,49.95) .. controls (225.62,49.89) and (223.26,52.19) .. (223.19,56.86) -- (222.79,86.38) .. controls (222.7,93.05) and (220.32,96.35) .. (215.65,96.28) .. controls (220.32,96.35) and (222.6,99.71) .. (222.51,106.38)(222.55,103.38) -- (222.1,135.9) .. controls (222.04,140.57) and (224.34,142.93) .. (229.01,143) ;
\draw   (427,302) .. controls (431.67,302) and (434,299.67) .. (434,295) -- (434,294) .. controls (434,287.33) and (436.33,284) .. (441,284) .. controls (436.33,284) and (434,280.67) .. (434,274)(434,277) -- (434,273) .. controls (434,268.33) and (431.67,266) .. (427,266) ;
\draw    (393.29,266.49) -- (418.29,266.49) ;
\draw    (240.29,210.49) -- (265.29,210.49) ;
\draw   (423,262) .. controls (427.67,261.95) and (429.98,259.6) .. (429.93,254.93) -- (429.09,165.43) .. controls (429.03,158.76) and (431.33,155.41) .. (436,155.37) .. controls (431.33,155.41) and (428.97,152.1) .. (428.9,145.43)(428.93,148.43) -- (428.06,55.93) .. controls (428.01,51.26) and (425.66,48.95) .. (420.99,49) ;
\draw   (231,148) .. controls (226.33,147.85) and (223.93,150.1) .. (223.78,154.77) -- (223.33,168.78) .. controls (223.11,175.44) and (220.67,178.7) .. (216.01,178.55) .. controls (220.67,178.7) and (222.89,182.1) .. (222.68,188.77)(222.78,185.77) -- (222.23,202.78) .. controls (222.08,207.44) and (224.33,209.85) .. (229,210) ;
\draw   (231,214) .. controls (226.33,214.01) and (224.01,216.35) .. (224.02,221.02) -- (224.09,242.43) .. controls (224.12,249.1) and (221.8,252.44) .. (217.13,252.45) .. controls (221.8,252.44) and (224.14,255.76) .. (224.16,262.43)(224.15,259.43) -- (224.27,295.04) .. controls (224.28,299.71) and (226.62,302.03) .. (231.29,302.02) ;
\draw    (241.14,302.75) -- (263.43,303.25) ;
\draw    (391.86,302.75) -- (414.14,303.25) ;
\draw  [dash pattern={on 0.84pt off 2.51pt}]  (242.29,238.49) -- (415,240) ;
\draw [fill={rgb, 255:red, 224; green, 18; blue, 43 }  ,fill opacity=1 ][line width=3]    (251.79,210.49) -- (252,240) ;
\draw [fill={rgb, 255:red, 224; green, 18; blue, 43 }  ,fill opacity=1 ][line width=3]    (401,238) -- (401,267) ;

\draw (197,78) node [anchor=north west][inner sep=0.75pt]   [align=left] {A};
\draw (448,274.4) node [anchor=north west][inner sep=0.75pt]    {$\emptyset $};
\draw (449,168) node [anchor=north west][inner sep=0.75pt]   [align=left] {B};
\draw (193,178) node [anchor=north west][inner sep=0.75pt]   [align=left] {B};
\draw (197,247.4) node [anchor=north west][inner sep=0.75pt]    {$\emptyset $};
\draw (240,22.4) node [anchor=north west][inner sep=0.75pt]    {$A\succ B$};
\draw (394,19.4) node [anchor=north west][inner sep=0.75pt]    {$B\succ A$};
\draw (269,133.4) node [anchor=north west][inner sep=0.75pt]    {$c_{A}$};
\draw (272,197.4) node [anchor=north west][inner sep=0.75pt]    {$c_{B} +\epsilon $};
\draw (341,253.4) node [anchor=north west][inner sep=0.75pt]    {$c_{B} -\epsilon '$};
\draw (271,288.4) node [anchor=north west][inner sep=0.75pt]    {$0$};
\draw (375,292.35) node [anchor=north west][inner sep=0.75pt]    {$0$};
\draw (268,41.35) node [anchor=north west][inner sep=0.75pt]    {$1$};
\draw (372,37.35) node [anchor=north west][inner sep=0.75pt]    {$1$};
\draw (320,215.4) node [anchor=north west][inner sep=0.75pt]    {$c_{B}$};

\end{tikzpicture}
    \caption{\footnotesize{An adjustment to the DA-STB cutoffs at school $B$. Students of type $A\succ B$ with lottery numbers in $[c_B,c_B+\epsilon)$ become unmatched, while students of type $B\succ A$ with lottery numbers in $[c_B-\epsilon',c_B)$ are newly assigned to $B$.}}
    \label{fig:stb-alt}
\end{figure}
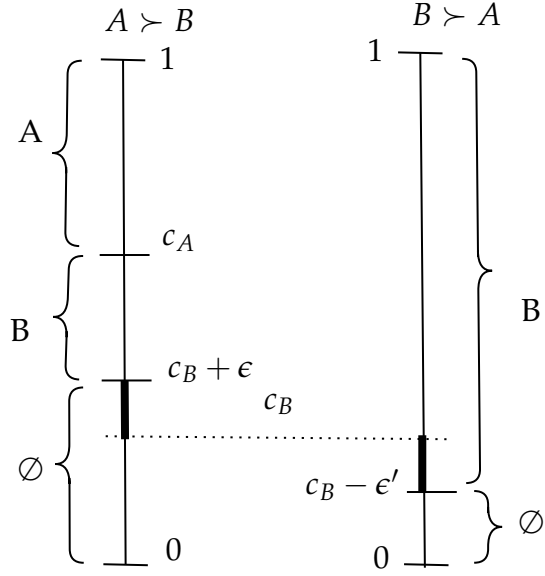

\paragraph{Adjustments to DA-STB.}
Every student assigned to $A$ under DA-STB ranks it first. In fact, among all full-capacity incentive-compatible assignments there is an optimum with this same property: no student who ranks $B$ first is assigned to $A$ (see Online Appendix~\ref{app:proof-stb-solution}). All $q_A$ seats at $A$ are then assigned to students who rank it first, which can be implemented keeping the cutoff at $A$ equal to $c_A$. We therefore keep $c_A^1=c_A^2=c_A$, raise the second-choice cutoff at $B$ to $c_B^2=c_B+\epsilon$, and lower its first-choice cutoff to $c_B^1=c_B-\epsilon'$, where $\epsilon,\epsilon'\geq0$. Figure~\ref{fig:stb-alt} illustrates the adjustment.

To preserve full capacity at $B$, the mass of students removed from $B$ must equal the mass newly admitted. Thus $\eta_A\epsilon=\eta_B\epsilon'$, or $\epsilon'=(\eta_A/\eta_B)\epsilon$. The change in the planner's payoff is then $\alpha\eta_B\epsilon'-(1-\alpha)\eta_A\epsilon=(2\alpha-1)\eta_A\epsilon$. Since $\alpha>1/2$, the planner wants to choose the largest feasible $\epsilon$.

There are three upper bounds on $\epsilon$. First, incentive compatibility for students of type $A\succ B$ requires $\epsilon\leq(1-c_A)\eta_B(u_1/u_2-1)$. The incentive constraint for students of type $B\succ A$ is strictly satisfied for all adjustments in this class.

Second, once the second-choice cutoff at $B$ reaches $c_A$, no student rejected from $A$ can be assigned to $B$. Raising this cutoff further has no effect on the assignment, so we may restrict attention to $\epsilon\leq c_A-c_B$.

Finally, the first-choice cutoff at $B$ must remain nonnegative: $c_B-\epsilon'\geq0$, or $\epsilon\leq(\eta_B/\eta_A)c_B$. The derivation of the incentive constraints is given in Online Appendix~\ref{app:proof-stb-solution}. We therefore obtain the following characterization.

\begin{proposition}\label{prop:stb-solution}
Suppose that $q_A+q_B<1$, that $\eta_A/q_A>\eta_B/q_B$, and that $\alpha>1/2$. Within the class of STB adjustments defined above, the optimal adjustment keeps the cutoff at $A$ unchanged and sets $c_B^2=c_B+\epsilon^*$ and $c_B^1=c_B-(\eta_A/\eta_B)\epsilon^*$, where
\begin{equation}
\epsilon^*
=
\min\left\{
(1-c_A)\eta_B\left(\frac{u_1}{u_2}-1\right),
\;c_A-c_B,
\;\frac{\eta_B}{\eta_A}c_B
\right\}.
\label{eq:stb-optimal-epsilon}
\end{equation}
\end{proposition}

\begin{proof}
See Online Appendix~\ref{app:proof-stb-solution}.
\end{proof}

\subsection{Comparison of MTB and STB adjustments}\label{sec:comparisonMTBSTB}

We now compare the optimal adjustments to DA-MTB and DA-STB. Under the conditions stated below, the two adjustment classes attain the same maximal mass of first-choice assignments. This conclusion differs from the comparison between the standard DA mechanisms: DA-STB may assign more students to their first choices than DA-MTB; see, for example, \citet{ashlagi2019assigning}.

\begin{proposition}\label{prop:comp}
Suppose that $q_A=q_B=q<\min\{\eta_A,\eta_B\}$ and that $\eta_A\geq\eta_B$. The optimal STB and MTB adjustments assign the same total mass of students to their first choices. If $\eta_A>\eta_B$, this mass is
\[
\begin{cases}
q+\eta_B q\left(1+\dfrac{u_1}{u_2}\right), & \text{if } \dfrac{u_1}{u_2}<\dfrac{\eta_A}{\eta_B},\\[1.2em] 2q,
& \text{if } \dfrac{u_1}{u_2}\geq\dfrac{\eta_A}{\eta_B}. \end{cases}
\]
If $\eta_A=\eta_B$, the common first-choice mass is $2q$.
\end{proposition}

\begin{proof}
See Appendix~\ref{compare}.
\end{proof}

The equality in Proposition~\ref{prop:comp} follows from the flexibility to condition cutoffs on reported rankings. Under the standard DA, the difference between STB and MTB is driven by rejection chains. Under MTB, a student rejected from one school has an independent lottery number at the next school and may displace a student who ranks that school first. Under STB, the same lottery number determines priority at both schools, so rejection from one school conveys information about priority at the other. This makes any displacement less likely. This explains the advantage of STB documented in the literature \citep{ashlagi2019assigning}.

With rank-dependent cutoffs, the planner can prevent this displacement. When $\eta_A>\eta_B$, the proof of Proposition~\ref{prop:comp} shows that the optimal MTB adjustment sets the second-choice cutoff at $A$ equal to one. Thus students who rank $A$ second are excluded from $A$, eliminating the displacement that puts MTB at a disadvantage under standard DA. If $\eta_A=\eta_B$, both classes of adjustments can assign every seat to a first-choice applicant.

The problem under either class of adjustments reduces to how much of $B$'s capacity can be reallocated from students who rank $A$ first to students who rank $B$ first, while preserving incentive compatibility. In terms of the mass of students who rank $B$ first and are assigned to $B$, the relevant incentive constraint is the same under STB and MTB.


\section{Implementation}\label{sec:implementation}

Our general formulation in Problem~\eqref{problem} involves choosing a feasible assignment. In practice, however, assignments must be operationalized in terms of admission cutoffs. Here we show that any feasible assignment can be represented through type-dependent cutoffs and priorities. If every school fills its capacity under \(x\), the cutoffs clear the market in the sense of \cite{azevedo2016supply}, and the corresponding realized cutoff allocation is stable under the constructed priorities and induces \(x\) ex ante.


\subsection{Cutoff representation with endogenous priorities}
\label{sec:priority-implementation}

We first illustrate the construction with an example.

\begin{example}\label{example:imple-MTB}
Consider two schools $A$ and $B$, with capacities $q_A=q_B=2/5$, and two preference types with masses $\eta_A=\eta_B=1/2$. Suppose that a feasible assignment $x$ gives the lotteries $p_{\succ_A}=(3/5,1/5,1/5)$ and $p_{\succ_B}=(1/5,3/5,1/5)$, where the coordinates correspond to $(A,B,\emptyset)$.

A student who ranks $A$ first receives $A$ with probability $3/5$, and is rejected from $A$ with probability $2/5$. Conditional on rejection from $A$, they must receive $B$ with probability $(1/5)/(2/5)=1/2$. Thus half of the students rejected from $A$ must receive $B$, and half must remain unmatched. Similarly, a student who ranks $B$ first receives $B$ with probability $3/5$ and, conditional on rejection from $B$, receives $A$ with probability $1/2$.

We construct cutoffs and report-dependent priority bonuses that reproduce these probabilities under MTB.\footnote{The same exercise can be carried out under STB. In that case, rejection from the first-choice school conveys information about the student's priority at the second-choice school, so the bonuses needed to reproduce the assignment differ from those under MTB. See Online Appendix, Section~\ref{example:imple-STB}.}
Each student draws independent lottery numbers $(\omega_A,\omega_B)\sim U[0,1]^2$. Let $c_A$ and $c_B$ be the school cutoffs. A student reporting $A\succ B$ receives priority bonuses $\rho_{AA}$ at $A$ and $\rho_{AB}$ at $B$, giving their scores $\omega_A+\rho_{AA}$ and $\omega_B+\rho_{AB}$. Similarly, a student reporting $B\succ A$ receives bonuses $\rho_{BA}$ at $A$ and $\rho_{BB}$ at $B$, giving their scores $\omega_A+\rho_{BA}$ and $\omega_B+\rho_{BB}$.

Only the difference between a cutoff and a priority bonus matters for eligibility. So we can normalize the second-choice bonuses to zero: $\rho_{AB}=\rho_{BA}=0$. The four remaining unknowns, $c_A,c_B,\rho_{AA}$, and $\rho_{BB}$, are determined by the assignment probabilities.

A student who ranks $A$ first must receive $A$ with probability $3/5$, so $\Pr(\omega_A+\rho_{AA}\geq c_A)=3/5$. Conditional on rejection from $A$, they must be eligible for $B$ with probability $1/2$. Since the lottery numbers are independent,
\[
\Pr\bigl(\omega_B\geq c_B
    \mid \omega_A+\rho_{AA}<c_A\bigr)
=
\Pr(\omega_B\geq c_B)
=
\frac12.
\]
Hence $c_B=1/2$. The same argument for students who rank $B$ first gives $c_A=1/2$.

With $c_A=1/2$, the first-choice assignment probability requires $\Pr(\omega_A+\rho_{AA}\geq1/2)=3/5$, so $\rho_{AA}=1/10$. Similarly, $\rho_{BB}=1/10$. Thus the assignment is reproduced by setting $c_A=c_B=1/2$, giving a bonus of $1/10$ to students who rank a school first, and giving no bonus to students who rank it second.$\blacksquare$
\end{example}

The following proposition extends this construction to a general environment.

\begin{proposition}\label{implementation}
Let $x$ be any feasible assignment. Under MTB, there exist school cutoffs $c_j\in[0,1]$ and report-dependent priority bonuses $\rho_j(\succ)\geq0$, for $(\succ,j)\in\mcR\times S$, such that the following rule induces $x$: each student draws independent school-specific lottery numbers $(\omega_j)_{j\in S}\sim U[0,1]^n$, independently of their report. If they report $\succ$, their priority score at school $j$ is $\omega_j+\rho_j(\succ)$. They receive their most preferred school, according to their report, among those for which $\omega_j+\rho_j(\succ)\geq c_j$, and remain unmatched if there is no such school. If every school fills its capacity under $x$, the cutoffs clear the market, the corresponding outcome is stable under the constructed priorities, and it induces \(x\) ex ante.
\end{proposition}

\begin{proof}
See Appendix~\ref{app:proof-implementation}.
\end{proof}

The bonuses in Proposition~\ref{implementation} depend on reported preferences but need not be monotone in the reported rank of a school. The construction is therefore more general than the rank-dependent adjustments studied in Section~\ref{sec:two-schools}.

Applied to a feasible assignment of Problem~\eqref{problem}, the construction reproduces the lottery associated with every admissible report. Since the assignment satisfies the incentive constraints of that problem, truthful reporting is optimal under the cutoff rule.

\subsection{Implementation through report-dependent priorities and acceptability}
\label{sec:reserve-implementation}

The implementation in Proposition~\ref{implementation} may not give $x$ through the student-proposing DA.  Example~\ref{example:no_imple} in Appendix~\ref{sec:example:no_imple} shows that this problem is not specific to our construction of priorities. However, we can always implement the assignment through the DA if schools may declare some students to be inadmissible. Let $\rho_j(\succ)$ and $c_j$ be the bonuses and cutoffs from Proposition~\ref{implementation}. School $j$ ranks students by $\omega_j+\rho_j(\succ)$ and finds a student acceptable if and only if $\omega_j+\rho_j(\succ)\geq c_j$. \textit{Students with scores below this cutoff are rejected even if the school has vacant seats.}

\begin{proposition}\label{prop:da-implementation}
Let $x\in X$ be any feasible assignment, and let $c_j$ and $\rho_j(\succ)$ be the cutoffs and bonuses constructed in Proposition~\ref{implementation}.  Suppose that each school $j$ ranks students by $\omega_j+\rho_j(\succ)$ and finds them acceptable if and only if their scores are at least $c_j$. Under the student-proposing DA, each report $\succ\in\mcR$ then induces the lottery $p_\succ(x)$ associated with $x$. If $x$ also satisfies the incentive constraints in Problem~\eqref{problem}, truthful reporting is optimal, and the truthful ex-ante DA assignment is $x$.
\end{proposition}

The proof of Proposition~\ref{prop:da-implementation} is in Appendix~\ref{app:proof-reserve-implementation}. The intuition is that each student has a target school: their most preferred school at which they are acceptable. By Proposition~\ref{implementation}, the mass of students whose target is school \(j\) is  \(\sum_{\succ\in\mcR}x_{\succ,j}\), which does not exceed \(q_j\). Hence, no acceptable student ever needs to be rejected because of capacity. All rejections during DA are due to unacceptability, so every student eventually reaches and is retained at their target school.

\section{Approximate Optimal Assignments} \label{Approx}

In many of our examples, the optimal matching is sparse in the sense that students of a given type are only assigned to a small number of schools. Here we show that this is a general feature of an approximation to a solution of Problem~\ref{problem}. An optimal assignment can be approximated by an assignment in which each  type is assigned positive mass to only \(O(\log|\mcR|/\epsilon^2)\) outcomes.
The approximation preserves not only the planner's objective and incentive compatibility, but also, for every preference type, the probability of receiving one of their top-\(k\) schools for every \(k\).

Our result is in terms of the cardinality of $\mcR$. With \(n\) schools, the set of strict preference orderings has cardinality \(n!\). In applications, however, only a much smaller collection of preference orderings is typically represented in the population. For example, San Francisco has $28$ high schools, and $28!$ vastly exceeds the world's population, let alone the school-age children in the city. But the reason why $\mcR$ is small goes beyond population size; rankings often reflect preference commonalities that limit the amount of heterogeneity present in the population. Here, we focus on  \(\mcR\subseteq\mathcal P\) of represented preference types in Problem~\ref{problem}, and study how the complexity of an approximately optimal assignment depends on \(|\mcR|\), rather than on the (typically) much larger set \(\mathcal P\).

Throughout this section, suppose that all represented preference types have the same mass, $\eta_\succ=\eta=\frac{1}{|\mcR|}$ for all $\succ\in \mcR$ and $\frac{q_j}{\eta}\in\mathbb N$ for all schools $j$.
Under these assumptions, the planner's objective can be written as
\[
\sum_{\succ\in\mcR}\sum_{j\in S}
v_\succ(j)x_{\succ,j}
=
\frac{1}{|\mcR|}
\sum_{\succ\in\mcR}v_\succ(x),
\]
and Problem~\eqref{problem} becomes
\begin{equation}
\label{new_prob}
\begin{aligned}
\max_{x\in X}
\quad&
\frac{1}{|\mcR|}
\sum_{\succ\in\mcR}v_\succ(x)
\\
\text{s.t.}\quad&
\frac{1}{\eta}
\sum_{j\in S}u_\succ(j)x_{\succ,j}
\geq
\frac{1}{\eta}
\sum_{j\in S}u_\succ(j)x_{\succ',j},
\qquad
\text{for all distinct }\succ,\succ'\in\mcR.
\end{aligned}
\end{equation}

We use the following notion of approximation.

\begin{definition}
\label{def:approx}
Let \(x^*\) be a solution to Problem~\eqref{new_prob}, and let \(\epsilon>0\).
A feasible assignment \(\widetilde x\in X\) is an \(\epsilon\)-approximate solution to Problem~\eqref{new_prob} if
\[
\sum_{\succ\in\mcR}\sum_{j\in S}
v_\succ(j)\widetilde x_{\succ,j}
\geq
\sum_{\succ\in\mcR}\sum_{j\in S}
v_\succ(j)x^*_{\succ,j}
-\epsilon,
\]
and
\[
\frac{1}{\eta}
\left[
\sum_{j\in S}u_\succ(j)\widetilde x_{\succ,j}
-
\sum_{j\in S}u_\succ(j)\widetilde x_{\succ',j}
\right]
\geq-\epsilon
\]
for all distinct \(\succ,\succ'\in\mcR\).
\end{definition}

For each \(\succ\in\mcR\) and \(k\in\{1,\ldots,n\}\), define
\[
F_\succ(k;x)
=
\frac{1}{\eta}
\sum_{\ell=1}^k
x_{\succ,\pi_\ell(\succ)}.
\]
Thus, \(F_\succ(k;x)\) is the probability that a student of type \(\succ\) is assigned to one of their top-\(k\) schools. We also define
$T_k(x) = \sum_{\succ\in\mcR}\eta F_\succ(k;x) = \sum_{\succ\in\mcR} \sum_{\ell=1}^k x_{\succ,\pi_\ell(\succ)}$, which is the total mass of students assigned to one of their top-\(k\) schools. The following is our approximation result.

\begin{theorem}\label{thm:approx}
Suppose that \(|\mcR|\geq 2\), and let \(x^*\) be a solution to Problem~\eqref{new_prob}. For every \(\epsilon>0\), there exists an \(\epsilon\)-approximate solution \(\widetilde x\in X\) such that each represented preference type is assigned with positive probability to at most
\[
1+\frac{4\ln|\mcR|}{\epsilon^2}
\]
outcomes in \(S\cup\{\emptyset\}\). In particular, if \(|\mcR|\) is polynomial in \(n\), the support of each represented preference type is $O\left(\frac{\ln n}{\epsilon^2}\right)$.

Moreover, for every \(\succ\in\mcR\) and every \(k\in\{1,\ldots,n\}\),
\[
\left|
F_\succ(k;\widetilde x)-F_\succ(k;x^*)
\right|
\leq\epsilon.
\]
Consequently,
\[
\left|T_k(\widetilde x)-T_k(x^*)\right|
\leq\epsilon
\qquad
\text{for every }k\in\{1,\ldots,n\}.
\]
\end{theorem}

Theorem~\ref{thm:approx} shows that the sparse assignment remains close to the optimum not only in terms of the planner's objective and incentive compatibility, but also in terms of the rank distribution of assignments faced by each preference type. In particular, the top-\(k\) conclusion shows that sparsification does not substantially alter the quality of assignments as measured by students' rankings. For every preference type, the probability of receiving one of the top \(k\) schools is preserved within \(\epsilon\), simultaneously for every \(k\). Consequently, the aggregate share \(T_k\) of students receiving a top-\(k\) assignment is also preserved within~\(\epsilon\).

The proof relies on the following Birkhoff--von Neumann-type property of the feasible-assignment polytope.

\begin{lemma}
\label{lem:bvn}
Every vertex \(m\) of \(X\) assigns each preference type deterministically to a single outcome in \(S\cup\{\emptyset\}\). Thus, $m_{\succ,j}\in\{0,\eta\}$ for all $\succ\in \mcR$ and $j\in S\cup\{\emptyset\}$; and each row of \(m\) contains exactly one nonzero entry.
\end{lemma}

The rest of the proof is an application of the probabilistic method. By showing that an $\epsilon$-approximate solution has a positive probability under a natural probabilistic model, we establish its existence. The proof of Lemma~\ref{lem:bvn} is in Appendix~\ref{proof:bvn}, and the proof of Theorem~\ref{thm:approx} is in Appendix~\ref{proof:approx}.

\section{Concluding Remarks}\label{sec:conclusion}

This paper studies how preference-dependent school priorities can be used while preserving incentives for truthful reporting. Our main result provides a characterization of when deferred acceptance under multiple tie-breaking (DA-MTB) solves the planner's problem. Under interior cutoffs, DA-MTB is optimal if and only if, for each school, the planner's payoff from assigning a student to that school is independent of the student's preference type. Thus, whenever the planner values assigning a given school differently across represented preference types, there is scope to improve upon DA-MTB while preserving incentive compatibility. Building on this observation, in the two-school environment, we characterize the optimal rank-dependent cutoff adjustments within the MTB and STB adjustment classes when the planner places greater weight on first-choice assignments. Under the conditions of our comparison result, the two optimized adjustment classes attain the same total mass of first-choice assignments.

We then study the representation and implementation of general feasible assignments. Allowing school priorities to depend on submitted preferences yields a cutoff representation of any feasible assignment under multiple tie-breaking. Report-dependent relative priorities alone, however, need not be sufficient for implementation through student-proposing DA. Once schools are allowed to deem some students unacceptable, the target report-contingent lotteries can be reproduced through DA, so that incentive-compatible assignments are truthfully implemented.

Finally, we show that under equal type masses and an integrality condition on capacities, the planner's optimum can be approximated by a feasible assignment that is approximately incentive compatible and has small support for each represented preference type. Moreover, the approximation preserves, up to a small error, the probability that each type receives one of their top-\(k\) schools, simultaneously for every \(k\).

\bibliographystyle{econometrica}
\bibliography{references}

\section{Appendix}

\subsection{Proof of Lemma \ref{lem:da-strict-ic}}\label{DA:proof}

\begin{proof}
Fix a preference ordering \(\succ\in\mcR\) and an alternative report \(\succ'\neq \succ\). Since an individual student has zero mass, a unilateral change in their report does not affect the DA-MTB cutoff vector.

For a lottery realization $\omega=(\omega_1,\ldots,\omega_n)\in[0,1]^n$, define the student's eligibility set by
\[ E(\omega) = \left\{ s\in S: \omega_s\geq c_s^{MTB} \right\}.
\]
This set depends on the lottery realization and the cutoff vector, but not on
the student's report.

If \(E(\omega)\neq\varnothing\), reporting an ordering \(r\) assigns the
student to the \(r\)-most-preferred school in \(E(\omega)\); if
\(E(\omega)=\varnothing\), they remain unmatched. Therefore, for every
\(\omega\), truthful reporting selects the \(\succ\)-most-preferred school
available in \(E(\omega)\). Hence
\[
u_\succ\bigl(\mu(\succ,\omega)\bigr)
\geq
u_\succ\bigl(\mu(\succ',\omega)\bigr)
\qquad
\text{for every }\omega\in[0,1]^n,
\]
where \(\mu(r,\omega)\) denotes the assignment obtained from report \(r\)
and lottery realization \(\omega\).

It remains to show that the inequality is strict on an event of positive
probability. Since \(\succ\neq\succ'\) and both are strict orderings, there
exist schools \(a,b\in S\) such that $a\succ b$ and $b\succ' a$.
Consider the event
\[
A_{a,b} = \left\{ \omega_a\geq c_a^{MTB}, \; \omega_b\geq c_b^{MTB}, \;
\omega_s<c_s^{MTB}
\text{ for every }s\in S\setminus\{a,b\}
\right\}.
\]
On \(A_{a,b}\), the eligibility set is exactly $E(\omega)=\{a,b\}$. Truthful reporting therefore assigns the student to \(a\), whereas reporting
\(\succ'\) assigns them to \(b\). Since \(a\succ b\), $u_\succ(a)>u_\succ(b)$.

Because the lottery numbers are independent across schools and every
DA-MTB cutoff is interior,
\[
\Pr(A_{a,b})
=
(1-c_a^{MTB})(1-c_b^{MTB})
\prod_{s\in S\setminus\{a,b\}}c_s^{MTB}
>0.
\]
Thus truthful reporting yields weakly higher utility for every lottery
realization and strictly higher utility on an event of positive probability.
Taking expectations gives
\[
\sum_{j\in S}
u_\succ(j)p^{MTB}_{\succ,j}
>
\sum_{j\in S}
u_\succ(j)p^{MTB}_{\succ',j}.
\]
\end{proof}

\subsection{Proof of Theorem \ref{thm:da-optimal}}\label{proof:cor:da}

\begin{proof}
Let \(x^{MTB}\) denote the DA-MTB assignment.

Suppose first that there exist a school \(j\) and two types
\(\succ,\succ'\in \mcR\) such that $$v_\succ(j)>v_{\succ'}(j)$$.
Because all DA-MTB cutoffs are interior, $x^{MTB}_{\succ,\emptyset}>0$ and $x^{MTB}_{\succ',j}>0$.

For sufficiently small \(\delta>0\), consider the adjustment that moves a mass \(\delta\) of type \(\succ\) from \(\emptyset\) to \(j\), and a mass \(\delta\) of type \(\succ'\) from \(j\) to \(\emptyset\). This preserves every type's total mass and leaves enrollment at every school unchanged, so the perturbed assignment is feasible.

The resulting change in the planner's payoff is
\[
\delta\bigl(v_\succ(j)-v_{\succ'}(j)\bigr)>0.
\]
By Lemma \ref{lem:da-strict-ic}, all incentive constraints are strict at \(x^{MTB}\). Since there are finitely many such constraints and they are linear in
the assignment, for sufficiently small \(\delta\) they remain satisfied after the adjustment. Hence \(x^{MTB}\) cannot be optimal for Problem \ref{problem}.

Conversely, suppose that for every school \(j\) there exists \(\bar v_j\) such that
\[
v_\succ(j)=\bar v_j
\qquad
\text{for every }\succ\in \mcR.
\]
Then every feasible assignment \(x\) satisfies
\[
\sum_{\succ\in R}\sum_{j\in S}v_\succ(j)x_{\succ,j}
=
\sum_{j\in S}\bar v_j\sum_{\succ\in R}x_{\succ,j}
\leq
\sum_{j\in S}\bar v_j q_j.
\]
Under interior DA-MTB cutoffs every school fills its capacity, so \(x^{MTB}\) attains this upper bound. Therefore, \(x^{MTB}\) is optimal.
\end{proof}

\subsection{Proof of \autoref{2schools}}\label{proof_2schools}


\begin{proof}
Since \(q_A,q_B>0\) and both schools fill, \(c_A,c_B<1\). If either \(c_A=0\) or \(c_B=0\), then \(\bar\epsilon=\bar\delta=0\), so \((\epsilon,\delta)=(0,0)\) is the only feasible pair. At this point both incentive constraints are strict, and case~(ii) follows immediately. Hence, for the remainder of the proof, we may assume \(c_A,c_B\in(0,1)\).

Because Problem (\ref{eq:planner-problem-2schools2}) restricts attention to capacity-preserving adjustments, both schools remain filled under every feasible MTB adjustment in the class. Subtracting the DA-MTB capacity equations from the corresponding capacity equations under the perturbed cutoffs gives
\begin{equation}\label{Prop2:proof1}
\eta_A\epsilon
=
\eta_B\left[
\epsilon'(c_B-\delta)+\delta(1-c_A)
\right],
\end{equation}

and
\begin{equation}\label{Prop2:proof2}
\eta_B\delta
=
\eta_A\left[
\epsilon(1-c_B)+\delta'(c_A-\epsilon)
\right].
\end{equation}

We first characterize the restrictions on \((\epsilon,\delta)\)
implied by the cutoff bounds. Since
\(0\leq\epsilon'\leq 1-c_A\),
\autoref{Prop2:proof1} can be satisfied if and only if
\[
\delta(1-c_A)
\leq
\frac{\eta_A}{\eta_B}\epsilon
\leq
(1-c_A)c_B.
\]
Similarly, since \(0\leq\delta'\leq 1-c_B\), \autoref{Prop2:proof2} can be satisfied if and only if
\[
\epsilon(1-c_B)
\leq
\frac{\eta_B}{\eta_A}\delta
\leq
(1-c_B)c_A.
\]

Therefore, the cutoff and capacity restrictions are equivalent to
\[
0\leq\epsilon\leq\bar\epsilon,
\qquad
0\leq\delta\leq\bar\delta,
\]
together with
\[
\frac{\eta_A}{\eta_B}(1-c_B)\epsilon
\leq
\delta
\leq
\frac{\eta_A}{\eta_B}
\frac{\epsilon}{1-c_A},
\]
where
\[
\bar\epsilon
=
\min\left\{
c_A,
\frac{\eta_B}{\eta_A}(1-c_A)c_B
\right\},
\qquad
\bar\delta
=
\min\left\{
c_B,
\frac{\eta_A}{\eta_B}(1-c_B)c_A
\right\}.
\]

We next verify that the northeast corner $(\bar\epsilon,\bar\delta)$ satisfies the last two inequalities. From the definition of
\(\bar\epsilon\),
\[
\frac{\eta_A}{\eta_B}(1-c_B)\bar\epsilon
\leq
\frac{\eta_A}{\eta_B}(1-c_B)c_A
\]
and
\[
\frac{\eta_A}{\eta_B}(1-c_B)\bar\epsilon
\leq
(1-c_A)(1-c_B)c_B
\leq c_B.
\]
Hence 

\begin{equation}\label{eq:bar_epsilon}
\frac{\eta_A}{\eta_B}(1-c_B)\bar\epsilon
\leq
\bar\delta.
\end{equation}

Similarly, from the definition of \(\bar\delta\),
\[
\frac{\eta_B}{\eta_A}(1-c_A)\bar\delta
\leq
\frac{\eta_B}{\eta_A}(1-c_A)c_B
\]
and
\[
\frac{\eta_B}{\eta_A}(1-c_A)\bar\delta
\leq
(1-c_A)(1-c_B)c_A
\leq c_A.
\]
Therefore,
\begin{equation}\label{eq:bar_delta}
\frac{\eta_B}{\eta_A}(1-c_A)\bar\delta
\leq
\bar\epsilon
\iff     \bar\delta
\leq
\frac{\eta_A}{\eta_B}
\frac{\bar\epsilon}{(1-c_A)}.
\end{equation}

Thus, absent the incentive constraints, the northeast corner
\((\bar\epsilon,\bar\delta)\) is feasible.

As we noted before, the objective is strictly
increasing in both \(\epsilon\) and \(\delta\).

Substituting the two capacity equations into the incentive constraints and using \(\eta_A+\eta_B=1\), the constraints reduce to 
\[
\delta
\leq
\frac{u_1}{u_2}
\frac{\eta_A}{\eta_B}\epsilon
+
(1-c_A)(1-c_B)
\left(
\frac{u_1}{u_2}-1
\right)\eta_A,
\qquad (IC_A)
\]
and
\[
\delta
\geq
\frac{u_2}{u_1}
\frac{\eta_A}{\eta_B}\epsilon
-
(1-c_A)(1-c_B)
\left(
1-\frac{u_2}{u_1}
\right)\eta_A.
\qquad (IC_B)
\]

Subtracting the $IC_B$
boundary from the $IC_A$ boundary gives
\[
\left(
\frac{u_1}{u_2}
-
\frac{u_2}{u_1}
\right)
\left[
\frac{\eta_A}{\eta_B}\epsilon
+
(1-c_A)(1-c_B)\eta_A
\right]
>0.
\]
Therefore, the $IC_A$ boundary lies strictly above the $IC_B$
boundary for every $\epsilon\geq 0$, and hence the two incentive
constraints cannot bind simultaneously.

We now distinguish three cases.

Suppose first that \(IC_A\) is violated at
\((\bar\epsilon,\bar\delta)\), or holds there with equality. The
largest value of \(\delta\) compatible with \(IC_A\) at
\(\epsilon=\bar\epsilon\) is
\[
\delta^*
=
\frac{u_1}{u_2}
\frac{\eta_A}{\eta_B}\bar\epsilon
+
(1-c_A)(1-c_B)
\left(
\frac{u_1}{u_2}-1
\right)\eta_A.
\]
By assumption, $\delta^*\leq\bar\delta$. Moreover,
\[
\delta^* > \frac{\eta_A}{\eta_B}\bar\epsilon > \frac{\eta_A}{\eta_B}(1-c_B)\bar\epsilon,
\]
while by \autoref{eq:bar_delta}
\[
\delta^* \leq\bar\delta \leq \frac{\eta_A}{\eta_B} \frac{\bar\epsilon}{1-c_A}.
\]
Hence \((\bar\epsilon,\delta^*)\) satisfies all the cutoff and capacity
restrictions. It also satisfies \(IC_B\) strictly.

Finally, because the right-hand side of \(IC_A\) is strictly increasing
in \(\epsilon\), every feasible pair \((\epsilon,\delta)\) satisfies $\epsilon\leq\bar\epsilon$ and
\[
\delta \leq \frac{u_1}{u_2} \frac{\eta_A}{\eta_B}\epsilon +
(1-c_A)(1-c_B)
\left(
\frac{u_1}{u_2}-1
\right)\eta_A
\leq
\delta^*.
\]
Thus \((\bar\epsilon,\delta^*)\) weakly dominates every feasible pair
coordinatewise. Since the objective is strictly increasing in both
coordinates, it is optimal. Therefore, $\epsilon^*=\bar\epsilon$, 
\(IC_A\) binds, and \(IC_B\) is slack. This is case~(i).

Suppose next that both incentive constraints are satisfied strictly at \((\bar\epsilon,\bar\delta)\). Since this point is feasible and the objective is strictly increasing in both coordinates, it is optimal: $\epsilon^*=\bar\epsilon$, $\delta^*=\bar\delta$. This is case~(ii).

Finally, the case when \(IC_B\) is violated at
\((\bar\epsilon,\bar\delta)\), or holds there with equality, is analogous to the first case analyzed and we include it in the Online Appendix, Section \ref{OA_Proof2}.

\end{proof}

\subsection{Proof of \autoref{prop:comp}}\label{compare}

\begin{proof}
Suppose first that $\eta_A>\eta_B$. We shall show that under either optimal adjustment, all seats at $A$ are assigned to students who rank it first. We then compare how many students who rank $B$ first are assigned to $B$.

For STB, the first claim follows from Proposition~\ref{prop:stb-solution}. The DA-STB cutoffs satisfy $c_A>c_B$, and the adjustment leaves the cutoff at $A$ unchanged while weakly lowering the first-choice cutoff at $B$. Any student who ranks $B$ first and is eligible for $A$ is therefore also eligible for $B$. So no such student is assigned to $A$, and all $q$ seats at $A$ go to students who rank it first.

We next establish the same property for MTB. In this part of the proof, $c_A$ and $c_B$ denote the DA-MTB cutoffs. The capacity conditions are $q=(1-c_A)(\eta_A+\eta_Bc_B)$ and $q=(1-c_B)(\eta_B+\eta_Ac_A)$. Since $q<\eta_A$, the first condition implies $\eta_B(1-c_A)c_B<\eta_Ac_A$. Similarly, $q<\eta_B$ implies $\eta_A(1-c_B)c_A<\eta_Bc_B$. Thus
\[
\bar\epsilon=\frac{\eta_B}{\eta_A}(1-c_A)c_B<c_A,
\qquad
\bar\delta=\frac{\eta_A}{\eta_B}(1-c_B)c_A<c_B.
\]

We claim that $IC_B$ is strictly satisfied at $(\bar\epsilon,\bar\delta)$. Since both schools have capacity $q$, equating the two capacity expressions and using $\eta_A+\eta_B=1$ gives $c_A-c_B=(\eta_A-\eta_B)(1-c_A)(1-c_B)$. Hence $c_A>c_B$, and
\[
\bar\delta
=\frac{\eta_A}{\eta_B}(1-c_B)c_A
>(1-c_B)c_A
>(1-c_A)c_B
=\frac{\eta_A}{\eta_B}\bar\epsilon.
\]
On the other hand, $IC_B$ requires
\[
\delta\geq
\frac{u_2}{u_1}\frac{\eta_A}{\eta_B}\epsilon
-(1-c_A)(1-c_B)
\left(1-\frac{u_2}{u_1}\right)\eta_A.
\]
Since $u_2/u_1<1$, the right-hand side at $\epsilon=\bar\epsilon$ is strictly smaller than $(\eta_A/\eta_B)\bar\epsilon$, which is smaller than $\bar\delta$. This proves the claim.

If $IC_A$ is also satisfied at $(\bar\epsilon,\bar\delta)$, this corner is optimal. Otherwise, case~(i) of Proposition~\ref{2schools} applies: $IC_A$ binds, $IC_B$ is slack, and $\epsilon$ is maximal. In either case, $\epsilon^*=\bar\epsilon$.

The capacity condition at $A$ under the optimal MTB adjustment is
\[
\eta_A\epsilon^*
=
\eta_B\left[
\epsilon^{\prime *}(c_B-\delta^*)
+\delta^*(1-c_A)
\right].
\]
Substituting $\epsilon^*=(\eta_B/\eta_A)(1-c_A)c_B$ and rearranging gives $(c_B-\delta^*)[(1-c_A)-\epsilon^{\prime *}]=0$. Since $\delta^*\leq\bar\delta<c_B$, we obtain $\epsilon^{\prime *}=1-c_A$. The second-choice cutoff at $A$ is therefore $c_A+\epsilon^{\prime *}=1$. Thus no positive mass of students reporting $B\succ A$ is assigned to $A$. Under truthful reporting, all $q$ seats at $A$ go to students who rank it first, as claimed.

Let $y$ be the mass of students who rank $B$ first and are assigned to $B$. Since both schools fill, the matched part of the assignment under either optimal adjustment is
\[
\begin{array}{c|cc}
 & A & B \\ \hline
A\succ B & q & q-y\\
B\succ A & 0 & y.
\end{array}
\]
The total mass of first-choice assignments is $q+y$. The planner's payoff is $\alpha(q+y)+(1-\alpha)(q-y)=q+(2\alpha-1)y$, so maximizing it is equivalent to maximizing $y$.

An individual report does not affect aggregate cutoffs in the continuum economy. Under either adjustment, a student reporting $A\succ B$ therefore receives the lottery given by the first row, divided by $\eta_A$, while a student reporting $B\succ A$ receives the lottery given by the second row, divided by $\eta_B$.

A student who ranks $A$ first obtains expected utility $(q/\eta_A)u_1+((q-y)/\eta_A)u_2$ from truthful reporting and $(y/\eta_B)u_2$ from reporting $B\succ A$. Thus $IC_A$ requires
\[
\frac{q}{\eta_A}u_1+\frac{q-y}{\eta_A}u_2
\geq\frac{y}{\eta_B}u_2.
\]
Multiplying by $\eta_A/u_2$ and using $\eta_A+\eta_B=1$ gives $y\leq\eta_Bq(1+u_1/u_2)$. Capacity at $B$ also requires $y\leq q$. Hence
\begin{equation}\label{y0}
y\leq
\min\left\{
q,\;\eta_Bq\left(1+\frac{u_1}{u_2}\right)
\right\}.
\end{equation}
We shall show that both optimal adjustments attain this bound.

Consider first MTB. We have shown that $\epsilon^*=\bar\epsilon$ and that the second-choice cutoff at $A$ is one. Since $IC_B$ is slack, Proposition~\ref{2schools} leaves two possibilities.

If $IC_A$ is satisfied at $(\bar\epsilon,\bar\delta)$, this corner is optimal and $\delta^*=\bar\delta=(\eta_A/\eta_B)(1-c_B)c_A$. The capacity condition at $B$ is
\[
\eta_B\delta^*
=
\eta_A\left[
\epsilon^*(1-c_B)
+\delta^{\prime *}(c_A-\epsilon^*)
\right].
\]
Substituting for $\delta^*$ gives $(c_A-\epsilon^*)[(1-c_B)-\delta^{\prime *}]=0$. Since $\epsilon^*=\bar\epsilon<c_A$, we obtain $\delta^{\prime *}=1-c_B$. The second-choice cutoff at $B$ is therefore one, so all $q$ seats at $B$ go to students who rank it first. Thus $y=q$, and incentive compatibility implies $q\leq\eta_Bq(1+u_1/u_2)$.

Otherwise, $IC_A$ binds, so $y=\eta_Bq(1+u_1/u_2)$. Feasibility then requires $\eta_Bq(1+u_1/u_2)\leq q$. Thus the optimal MTB adjustment attains the bound in \eqref{y0} in either case.

Now consider STB, and let $c_A$ and $c_B$ denote the DA-STB cutoffs. These are $c_A=1-q/\eta_A$ and $c_B=1-2q$. By Proposition~\ref{prop:stb-solution}, the optimal adjustment leaves the cutoff at $A$ unchanged. Students who rank $B$ first face cutoff $c_B-\epsilon'$, so $y=\eta_B(1-c_B+\epsilon')$. Since capacity preservation requires $\eta_B\epsilon'=\eta_A\epsilon$, we have
\begin{equation}\label{y2}
y=2\eta_Bq+\eta_A\epsilon.
\end{equation}

The bound $\epsilon\leq(\eta_B/\eta_A)c_B$ in Proposition~\ref{prop:stb-solution} cannot bind. Indeed, $c_A-c_B=q(2\eta_A-1)/\eta_A$, and $c_A-c_B<(\eta_B/\eta_A)c_B$ is equivalent to $q<\eta_B$, which holds by assumption. Thus $\epsilon^*$ is determined by either the incentive constraint or the bound $c_A-c_B$.

If the incentive constraint binds, then $\epsilon^*=(q\eta_B/\eta_A)(u_1/u_2-1)$. Substituting into \eqref{y2} gives $y=\eta_Bq(1+u_1/u_2)$.

If instead $\epsilon^*=c_A-c_B$, the second-choice cutoff at $B$ equals $c_A$. A student rejected from $A$ has a lottery number below $c_A$ and therefore cannot meet this cutoff at $B$. No positive mass of students who rank $A$ first is then assigned to $B$. Since $B$ fills, $y=q$.

Thus the optimal STB adjustment also attains the bound in \eqref{y0}. Both classes therefore assign the same mass $q+\min\{q,\eta_Bq(1+u_1/u_2)\}$ to first choices. Since $\eta_Bq(1+u_1/u_2)\geq q$ if and only if $u_1/u_2\geq\eta_A/\eta_B$, this mass is
\[
\begin{cases}
q+\eta_Bq\left(1+\dfrac{u_1}{u_2}\right),
& \text{if } \dfrac{u_1}{u_2}<\dfrac{\eta_A}{\eta_B},\\[1.2em]
2q,
& \text{if } \dfrac{u_1}{u_2}\geq\dfrac{\eta_A}{\eta_B}.
\end{cases}
\]

Finally, suppose that $\eta_A=\eta_B=1/2$. Under DA-STB, the cutoffs coincide at $c_A=c_B=1-2q$. A student is therefore eligible for both schools or neither, and every eligible student receives their first choice. Each school receives mass $(1/2)(2q)=q$. Thus DA-STB itself, which belongs to the class of STB adjustments, assigns all $2q$ seats to first-choice applicants and is optimal within this class.

Under DA-MTB, symmetry gives a common cutoff $c$ satisfying $q=(1-c^2)/2$, so $c=\sqrt{1-2q}$. Consider the rank-dependent cutoffs $c_A^1=c_B^1=1-2q$ and $c_A^2=c_B^2=1$. Since $q\in(0,1/2)$, we have $0<1-2q<\sqrt{1-2q}<1$, so these cutoffs belong to the class of MTB adjustments studied here.

Each type receives its first choice with probability $2q$ and its second choice with probability zero. Each school therefore receives mass $(1/2)(2q)=q$, so both capacities are filled. Truthful reporting gives expected utility $2qu_1$, while reversing the ranking gives $2qu_2$. Since $u_1>u_2$, truthful reporting is strictly optimal. Thus the MTB class also attains the maximal first-choice mass $2q$.
\end{proof}

\subsection{Proof of Proposition~\ref{implementation}}
\label{app:proof-implementation}

\begin{proof}
Let $p_{\succ,j}=x_{\succ,j}/\eta_\succ$ be the probability that a student reporting $\succ$ receives school $j$ under $x$. For each report $\succ\in\mcR$, write $\pi_1(\succ)\succ\cdots\succ\pi_n(\succ)$ for the schools in order of preference.

We first calculate the probability of admission to each school conditional on rejection from all preferred schools. Denote this probability by $a_j(\succ)$. At the first-choice school, we need $a_{\pi_1(\succ)}(\succ)=p_{\succ,\pi_1(\succ)}$. The student reaches their second choice with probability $1-p_{\succ,\pi_1(\succ)}$, so the required conditional admission probability is $a_{\pi_2(\succ)}(\succ)=p_{\succ,\pi_2(\succ)}/(1-p_{\succ,\pi_1(\succ)})$, provided the denominator is positive.

More generally, define
\begin{equation}
a_{\pi_k(\succ)}(\succ)
=
\frac{p_{\succ,\pi_k(\succ)}}
{1-\sum_{\ell=1}^{k-1}p_{\succ,\pi_\ell(\succ)}}
\label{eq:conditional-admission}
\end{equation}
whenever the denominator is positive. If it is zero, the student receives one of their first $k-1$ choices with probability one, so $p_{\succ,\pi_k(\succ)}=0$ and they never reach their $k$-th choice. The conditional admission probability then has no effect on the assignment, and we set $a_{\pi_k(\succ)}(\succ)=0$. Feasibility implies that $p_{\succ,\pi_k(\succ)}\leq1-\sum_{\ell=1}^{k-1}p_{\succ,\pi_\ell(\succ)}$, so all these admission probabilities belong to $[0,1]$.

We now choose school cutoffs and priority bonuses to reproduce these probabilities. For each school $j$, we want $\Pr(\omega_j+\rho_j(\succ)\geq c_j)=a_j(\succ)$ for every report $\succ$. Since $\omega_j$ is uniform on $[0,1]$, we can obtain these probabilities by choosing $1-c_j+\rho_j(\succ)=a_j(\succ)$.

Only the difference between a cutoff and a bonus matters, so we normalize one bonus at each school to zero. Fix a school $j$, choose $\widehat{\succ}\in\arg\min_{\succ\in\mcR}a_j(\succ)$, and set $\rho_j(\widehat{\succ})=0$. The cutoff and remaining bonuses are then
\[
c_j=1-a_j(\widehat{\succ}),
\qquad
\rho_j(\succ)=a_j(\succ)-a_j(\widehat{\succ}).
\]
Thus $c_j\in[0,1]$, and all bonuses are nonnegative. Repeat this construction for every school.

These choices make $\omega_j+\rho_j(\succ)\geq c_j$ equivalent to $\omega_j\geq1-a_j(\succ)$. Hence the probability that a student reporting $\succ$ is eligible for $j$ is exactly $a_j(\succ)$.

We verify by induction that the resulting assignment reproduces $x$. Fix a report $\succ$. The student is eligible for their first-choice school with probability $a_{\pi_1(\succ)}(\succ)=p_{\succ,\pi_1(\succ)}$, which is therefore their probability of receiving that school.

Suppose the construction reproduces the assignment probabilities for their first $k-1$ choices. They then reach their $k$-th choice with probability $1-\sum_{\ell=1}^{k-1}p_{\succ,\pi_\ell(\succ)}$. Under MTB, the lottery number at this school is independent of the lottery numbers at all preferred schools. Conditional on rejection from those schools, their probability of eligibility therefore remains $a_{\pi_k(\succ)}(\succ)$. If the probability of reaching the school is positive, their assignment probability is
\[
\left(1-\sum_{\ell=1}^{k-1}p_{\succ,\pi_\ell(\succ)}\right)
a_{\pi_k(\succ)}(\succ)
=
p_{\succ,\pi_k(\succ)},
\]
by \eqref{eq:conditional-admission}. If it is zero, both the target and induced assignment probabilities are zero.

The construction therefore assigns a student reporting $\succ$ to each school $j$ with probability $p_{\succ,j}$. Their probability of remaining unmatched is consequently $1-\sum_{j\in S}p_{\succ,j}=p_{\succ,\emptyset}$.

Under truthful reporting, a mass $\eta_\succ$ submits report $\succ$, so the mass assigned to $j$ is $\eta_\succ p_{\succ,j}=x_{\succ,j}$. The ex ante assignment is therefore $x$. Since $x$ is feasible, $\sum_{\succ\in\mcR}x_{\succ,j}\leq q_j$ for every school $j$, so the induced assignment respects school capacities. If every school fills under $x$, these inequalities hold with equality, and the constructed cutoffs clear the market.
\end{proof}

\subsection{Example of Section \ref{sec:reserve-implementation}}
\label{sec:example:no_imple} 

\begin{example}\label{example:no_imple} 
Consider four schools \(A,B,C,D\), each with capacity $q_A=q_B=q_C=q_D=\frac14$ and four equally represented preference types, $\eta_{\succ_1} = \eta_{\succ_2} = \eta_{\succ_3} = \eta_{\succ_4} = \frac14$, 
with
\[
\begin{aligned}
\succ_1 &: A\succ C\succ D\succ B,\\
\succ_2 &: B\succ D\succ A\succ C,\\
\succ_3 &: C\succ D\succ A\succ B,\\
\succ_4 &: B\succ A\succ D\succ C.
\end{aligned}
\]
Suppose student utilities depend only on rank and satisfy
\[
u_1=1,\qquad
u_2=\frac{19}{20},\qquad
u_3=\frac12,\qquad
u_4=\frac1{20},
\]
while the planner's rank-dependent payoffs are
\[
v_1=1,\qquad
v_2=\frac45,\qquad
v_3=\frac12,\qquad
v_4=\frac3{10}.
\]

Solving the linear program in Problem~\eqref{problem} gives the following optimal normalized assignment: 
\begin{equation}
P^*
=
\begin{pmatrix}
1&0&0&0\\[1mm]
0&\frac1{20}&0&\frac{19}{20}\\[1mm]
0&0&\frac{19}{20}&\frac1{20}\\[1mm]
0&\frac{19}{20}&\frac1{20}&0
\end{pmatrix},
\label{eq:target-P}
\end{equation}
where the columns correspond to \(A,B,C,D\).

\textbf{We claim that no profile of school priorities can implement \(P^*\) through student-proposing DA.}\footnote{The role of school \(A\) and type \(\succ_1\) is auxiliary. Type \(\succ_1\) exactly fills \(A\), ensuring that any \(\succ_4\)-student rejected from \(B\) is also rejected from \(A\) and therefore proceeds to \(D\). At the same time, the position of \(A\) distinguishes the reports \(\succ_2\) and \(\succ_4\), which would otherwise coincide once the auxiliary school is removed.}

Suppose otherwise. In the first round all \(\succ_1\)-students apply to
\(A\). Their mass is exactly \(q_A=1/4\), and \(P^*\) assigns all of them
to \(A\). Hence \(A\) cannot reject any of these students.

At \(B\), all \(\succ_2\)- and \(\succ_4\)-students apply in the first
round. In order to reproduce \(P^*\), \(B\) must reject a fraction
\(19/20\) of type \(\succ_2\) and a fraction \(1/20\) of type
\(\succ_4\). 

Under the target assignment, the \(\succ_2\)-students rejected from \(B\) eventually apply to \(D\), while the \(\succ_4\)-students rejected from \(B\) apply first to \(A\), are rejected there, and then apply to \(D\). Their total mass is
\[
\frac14\frac{19}{20}+\frac14\frac1{20}=\frac14=q_D.
\]
Hence, before any \(\succ_3\)-student reaches \(D\), school \(D\) never faces more than its capacity in applicants from types \(\succ_2\) and \(\succ_4\). Under universal school-side acceptability, it therefore cannot reject any of them.

For a \(\succ_3\)-student to reach \(D\), they must first be rejected from \(C\). But all \(\succ_3\)-students initially apply to \(C\), and their total mass is exactly \(q_C=1/4\). Thus \(C\) cannot reject any \(\succ_3\)-student unless another student first applies to \(C\). Under the target assignment, the only such students are the \(\succ_4\)-students who must eventually be assigned to \(C\). Yet a \(\succ_4\)-student reaches \(C\) only after being rejected from \(D\).

Thus a rejection of a \(\succ_4\)-student from \(D\) requires a prior arrival of a \(\succ_3\)-student at \(D\), while an arrival of a \(\succ_3\)-student at \(D\) requires a prior arrival of a \(\succ_4\)-student at \(C\). The rejection cycle has no event that can initiate it. Therefore student-proposing DA cannot generate \(P^*\) under universal school-side acceptability, regardless of the schools' relative priority orderings.$\blacksquare$
\end{example}

\subsection{Proof of Proposition~\ref{prop:da-implementation}}
\label{app:proof-reserve-implementation}

\begin{proof}
Write $s_j(\succ,\omega_j)=\omega_j+\rho_j(\succ)$ for a student's priority score at school $j$, and $r_j=c_j$ for the school's acceptability threshold.

Fix the truthful report profile and the lottery realizations. For each student, define their \textbf{target school} as their most preferred acceptable school, where acceptability is determined by
\[
s_j(\succ,\omega_j)\geq r_j=c_j.
\]
If no school is acceptable, their target is \(\emptyset\). By Proposition~\ref{implementation}, the resulting target assignment induces \(x\).

For each school \(j\), let \(T_j\) denote the set of students whose target school is \(j\). Since the target assignment is \(x\),
\[
\operatorname{mass}(T_j)=\sum_{\succ\in\mcR}x_{\succ,j}\leq q_j.
\]

We now show inductively that every rejection during student-proposing DA is due to unacceptability and that no acceptable student is ever rejected because of capacity.

In the first round, a student applies to their first-choice school. If they are unacceptable there, the school rejects them. If they are acceptable, then their first choice is their most preferred acceptable school, so they belong to the corresponding target set \(T_j\). Since \(\operatorname{mass}(T_j)\leq q_j\), the school can retain all such acceptable applicants.

Suppose the claim holds through round \(k-1\), and consider a student who reaches their \(k\)-th choice \(j=\pi_k(\succ)\). By the induction hypothesis, they were rejected from every school ranked above \(j\) because they were unacceptable there. If they are also unacceptable at \(j\), they are rejected and proceed to their next choice. If they are acceptable at \(j\), then \(j\) is their most preferred acceptable school, so they belong to \(T_j\). At any stage of the algorithm, the acceptable applicants retained by \(j\) are therefore a subset of \(T_j\). Since
\[
\operatorname{mass}(T_j)\leq q_j,
\]
no acceptable applicant is ever rejected from \(j\) because of capacity.

It follows that every student is rejected from exactly the schools they rank above their target school and they are then retained at their target school. A student with no acceptable school is rejected by every school and remains unmatched. Hence, under truthful reporting, DA generates the same allocation as the cutoff rule in Proposition~\ref{implementation}, and its ex ante assignment is \(x\).

It remains to verify incentives. Because a single student has zero mass, a unilateral change in their report does not affect aggregate capacities or the allocation of the other students. If they submit some report \(\succ'\), the same argument implies that they are assigned to their most preferred school, according to \(\succ'\), at which
\[
s_j(\succ',\omega_j)\geq r_j.
\]
Thus report \(\succ'\) induces exactly the cutoff-rule lottery \(p_{\succ'}\) constructed in Proposition~\ref{implementation}. If \(x\) also satisfies the incentive-compatibility constraints of Problem~\eqref{problem}, its report-contingent lottery menu satisfies
\[
\sum_{j\in S}u_\succ(j)p_{\succ,j}
\geq
\sum_{j\in S}u_\succ(j)p_{\succ',j}
\qquad\text{for every }\succ,\succ'\in\mcR.
\]
Therefore, truthful reporting is optimal. Hence student-proposing DA with the constructed priorities and acceptability thresholds implements \(x\).
\end{proof}

\subsection{Proof of Lemma~\ref{lem:bvn}}
\label{proof:bvn}

\begin{proof}
Let $Y:=\frac{1}{\eta}X = \left\{ \frac{1}{\eta}x:x\in X \right\}.$ If \(y\in Y\), then $\sum_{j\in S\cup\{\emptyset\}}y_{\succ,j}=1$ for any $\succ\in \mcR$; and $\sum_{\succ\in\mcR}y_{\succ,j} \leq \frac{q_j}{\eta}$ for any $j\in S$. Moreover, $y_{\succ,j} \geq 0$ for any $(\succ,j)\in\mcR\times(S\cup\{\emptyset\})$.

Since every row of \(y\) sums to one, $\sum_{\succ\in\mcR}y_{\succ,\emptyset} \leq |\mcR|$. This constraint is redundant, so adding it does not change the set \(Y\).

Consider now the bipartite graph $G= \bigl( \mcR,\, S\cup\{\emptyset\};\, E \bigr)$ with $E= \mcR\times(S\cup\{\emptyset\})$. Thus, each variable \(y_{\succ,j}\) corresponds to the edge joining the preference type \(\succ\) to the outcome \(j\).

Let \(A\) be the node-edge incidence matrix of \(G\).\footnote{The node-edge incidence matrix of a graph has one row for each node and one column for each edge. The entry corresponding to node \(v\) and edge \(e\) equals \(1\) if \(v\) is incident to \(e\), and equals \(0\) otherwise. Thus, for an undirected graph, each column contains exactly two entries equal to \(1\), corresponding to the two endpoints of the edge.} Hence \(A\) has $|\mcR|+n+1$ rows, one for each node of the graph, and $|\mcR|(n+1)$ columns, one for each variable \(y_{\succ,j}\). The column corresponding to \(y_{\succ,j}\) has a \(1\) in the row associated with \(\succ\), a \(1\) in the row associated with \(j\), and zeros elsewhere.

The constraints defining \(Y\) can therefore be written using the matrix \(A\): the rows corresponding to preference types are imposed as equalities,
\[
\sum_{j\in S\cup\{\emptyset\}}y_{\succ,j}=1,
\qquad
\succ\in\mcR,
\]
while the rows corresponding to schools and to the outside option are imposed as inequalities,
\[
\sum_{\succ\in\mcR}y_{\succ,j}
\leq
\frac{q_j}{\eta},
\qquad
j\in S,
\]
and
\[
\sum_{\succ\in\mcR}y_{\succ,\emptyset}
\leq
|\mcR|.
\]

Since \(G\) is bipartite, its node-edge incidence matrix \(A\) is totally unimodular; see, for example,
\citet[Chapter~19]{schrijver1986theory}.

The right-hand side of the system is integral. Indeed, the constraints associated with preference types have right-hand side \(1\), the outside-option constraint has right-hand side \(|\mcR|\), and, by assumption,
\[
\frac{q_j}{\eta}\in\mathbb N
\qquad
\text{for every }j\in S.
\]

By the Hoffman--Kruskal theorem, a polyhedron defined by a totally unimodular constraint matrix and an integral right-hand side has integral vertices; see, for example, \citet[Chapter~19]{schrijver1986theory}. Therefore every vertex of \(Y\) is integral.

Let \(y\) be a vertex of \(Y\). Since \(y\geq0\), all its entries are nonnegative integers. Moreover, for every \(\succ\in\mcR\),
\[
\sum_{j\in S\cup\{\emptyset\}}y_{\succ,j}=1.
\]
It follows that exactly one entry in each row of \(y\) is equal to \(1\), while all the remaining entries are equal to \(0\).
Finally, multiplication by the positive scalar \(\eta\) maps vertices of \(Y\) one-to-one onto vertices of \(X\). Hence every vertex \(m\) of \(X\) satisfies
\[
m_{\succ,j}\in\{0,\eta\},
\]
with exactly one nonzero entry in each row.
\end{proof}

\subsection{Proof of Theorem \ref{thm:approx}}\label{proof:approx}

\begin{proof}
Let \(x^*\) be a solution to the planner's problem. Let \(M\) be the set of the vertices
of \(X\). Write \(x^*\) as a convex combination of the matrices in \(M\). Interpret this
convex combination as a probability distribution over \(M\) and consider \(k\) i.i.d. draws
from this distribution:
\[
(\widetilde m_1,\ldots,\widetilde m_k).
\]
Calculate the sample average as
\[
\widetilde x=\frac{1}{k}\sum_{i=1}^k\widetilde m_i.
\]
Note that the expectation of \(\widetilde x\) is \(x^*\):
\[
\mathbb E(\widetilde x)
=
\frac{1}{k}\sum_{i=1}^k\mathbb E(\widetilde m_i)
=
\frac{1}{k}\sum_{i=1}^k x^*
=
x^*.
\]

\paragraph{First Step.}
We use McDiarmid's inequality to bound the violation of the IC constraints.
Define the matrix \(u_{\succ,\succ'}\) of dimension
\(|\mcR|\times(n+1)\) such that it has \(u_\succ\) in row \(\succ\),
\(-u_\succ\) in row \(\succ'\), and zeros in all other rows. Let:

\begin{enumerate}
\item \(X_i=\widetilde m_i\), \(i=1,\ldots,k\).
\item \(\mathcal X_i=M\), \(i=1,\ldots,k\).
\item
\[
f(m_1,\ldots,m_k)
=
u_{\succ,\succ'}\cdot
\frac{1}{\eta}
\left(
\frac{1}{k}
\right)
\sum_{i=1}^k m_i.
\]
\end{enumerate}

Consider two vectors of matrices $(m_1,\ldots,m_k)$ and $(m'_1,\ldots,m'_k)$, such that $m_j=m'_j$ for every $j\neq i$. Then
\begin{align*}
&
\left|
u_{\succ,\succ'}\cdot
\frac{1}{\eta}
\left(
\frac{1}{k}
\right)
\sum_{j=1}^k m_j
-
u_{\succ,\succ'}\cdot
\frac{1}{\eta}
\left(
\frac{1}{k}
\right)
\sum_{j=1}^k m'_j
\right|
\\
&=
\left|
u_{\succ,\succ'}\cdot
\frac{1}{\eta}
\left(
\frac{1}{k}
\right)
(m_i-m'_i)
\right|.
\end{align*}

Note that all the elements of each matrix
\(\frac{1}{\eta}m_i\) are either \(0\) or \(1\), and all the elements of
\(u_{\succ,\succ'}\) belong to \([-1,1]\). Hence,
\[
\left|
u_{\succ,\succ'}\cdot
\frac{1}{\eta}
\left(
\frac{1}{k}
\right)
(m_i-m'_i)
\right|
\leq
\frac{2}{k}.
\]
Therefore,
\[
|f(m)-f(m')|
\leq
\frac{2}{k}.
\]

Also,
\[
\mathbb E[
f(\widetilde m_1,\ldots,\widetilde m_k)
]
=
u_{\succ,\succ'}\cdot
\frac{1}{\eta}
x^*.
\]

Applying McDiarmid's inequality, we obtain
\[
\Pr\left(
u_{\succ,\succ'}\cdot\frac{1}{\eta}\widetilde x
-
u_{\succ,\succ'}\cdot\frac{1}{\eta}x^*
<
-\epsilon
\right)
\leq
\exp\left(
-\frac{k\epsilon^2}{2}
\right).
\]

Since \(x^*\) is incentive compatible,
\[
u_{\succ,\succ'}\cdot\frac{1}{\eta}x^*\geq0
\qquad
\text{for all distinct }\succ,\succ'\in\mcR.
\]
Hence
\[
\Pr\left(
u_{\succ,\succ'}\cdot\frac{1}{\eta}\widetilde x
<
-\epsilon
\right)
\leq
\exp\left(
-\frac{k\epsilon^2}{2}
\right).
\]

\paragraph{Second Step.}
We use McDiarmid's inequality to bound the deviation from the optimal value.

Consider the function \(g\), such that if
\(x=(m_1,\ldots,m_k)\),
\[
g(x)
=
\frac{1}{k}
\sum_{j=1}^k
\left(
\sum_{\succ\in\mcR}
\frac{1}{|\mcR|}
v_\succ(m_j)
\right).
\]

As before, take two vectors of matrices $x=(m_1,\ldots,m_k)$, and 
$x'=(m'_1,\ldots,m'_k),$ such that $m_j=m'_j$ for every $j\neq i$. Then
\begin{align*}
|g(x)-g(x')|
&=
\left|
\frac{1}{k|\mcR|}
\left(
\sum_{\succ\in\mcR}v_\succ(m_i)
-
\sum_{\succ\in\mcR}v_\succ(m'_i)
\right)
\right|.
\end{align*}

Since \(m_i\) and \(m'_i\) are two vertex matrices, and for every
\(\succ\in\mcR\), $0\leq v_\succ(m_i)\leq1$, and $0\leq v_\succ(m'_i)\leq1$,
we have
\[
\left|
\sum_{\succ\in\mcR}v_\succ(m_i)
-
\sum_{\succ\in\mcR}v_\succ(m'_i)
\right|
\leq
|\mcR|.
\]
Therefore,
\[
|g(x)-g(x')|
\leq
\frac{1}{k}.
\]

Moreover,
\[
\mathbb E[g(\widetilde m_1,\ldots,\widetilde m_k)]
=
\sum_{\succ\in\mcR}
\frac{1}{|\mcR|}
v_\succ(x^*).
\]

Applying McDiarmid's inequality,
\[
\Pr\left(
\sum_{\succ\in\mcR}
\frac{1}{|\mcR|}
v_\succ(\widetilde x)
-
\sum_{\succ\in\mcR}
\frac{1}{|\mcR|}
v_\succ(x^*)
<
-\epsilon
\right) \leq e^{-2k\epsilon^2}.
\]

\paragraph{Third Step.}
We now bound the deviation in the probabilities of receiving a top-\(k\) school.

Fix a preference type \(\succ\in\mcR\).
For each draw \(i=1,\ldots,k\), let $R_\succ^i\in\{1,\ldots,n+1\}$ denote the rank, according to \(\succ\), of the outcome assigned to type
\(\succ\) by \(\widetilde m_i\), where the outside option is assigned rank
\(n+1\).

By Lemma~\ref{lem:bvn}, each vertex assigns type \(\succ\) deterministically to
one outcome. Moreover, since the matrices \(\widetilde m_1,\ldots,\widetilde m_k\) are i.i.d., the random variables $R_\succ^1,\ldots,R_\succ^k$ are also i.i.d.

For every \(\ell\in\{1,\ldots,n\}\),
\[
\Pr(R_\succ^i\leq\ell) = \frac1\eta\sum_{r=1}^{\ell}\mathbb E[\widetilde m_{i,\succ,\pi_r(\succ)}] = F_\succ(\ell;x^*).
\]
Indeed, the probability distribution over vertices has mean \(x^*\).

On the other hand,
\[
F_\succ(\ell;\widetilde x)
=
\frac{1}{k}
\sum_{i=1}^k
\mathbf 1\{R_\succ^i\leq\ell\}.
\]
Thus, \(F_\succ(\cdot;\widetilde x)\) is the empirical distribution function of
the sampled ranks.

By the Dvoretzky--Kiefer--Wolfowitz--Massart inequality \citep{massart1990tight},
\[
\Pr\left(\max_{1\leq\ell\leq n} \left| F_\succ(\ell;\widetilde x) - F_\succ(\ell;x^*) \right| > \epsilon \right)
\leq 2e^{-2k\epsilon^2}.
\]

\paragraph{Final Step.}
Using the union bound, the probability that at least one IC constraint is violated by more than \(\epsilon\), that the planner's objective is more than \(\epsilon\) below the optimum, or that the top-\(\ell\) probability differs from that under \(x^*\) by more than \(\epsilon\) for some preference type, is bounded above by
\[
|\mcR|(|\mcR|-1)
\exp\left(
-\frac{k\epsilon^2}{2}
\right)
+
(2|\mcR|+1)\exp(-2k\epsilon^2).
\]

Let $C:=|\mcR|(|\mcR|-1)+1$. Choose $k=K_\epsilon := \left\lfloor \frac{2}{\epsilon^2} \ln C \right\rfloor +1.$ Then
\[
e^{\left( -\frac{k\epsilon^2}{2} \right)} < \frac{1}{C}, \quad \text{ and therefore }\quad e^{-2k\epsilon^2} < \frac{1}{C^4}.
\]
Hence the probability above is strictly smaller than $\frac{C-1}{C} + \frac{2|\mcR|+1}{C^4}$. Since \(|\mcR|\geq2\), we have $C\geq|\mcR|+1\geq3$, and therefore $C^3>2|\mcR|+1$. It follows that
\[
\frac{C-1}{C}
+
\frac{2|\mcR|+1}{C^4}
<1.
\]

Thus, with strictly positive probability, \(\widetilde x\) satisfies all IC
constraints up to \(\epsilon\), its objective value is within \(\epsilon\) of the
optimal value from below, and
\[
\max_{\succ\in\mcR}
\max_{1\leq\ell\leq n}
\left|
F_\succ(\ell;\widetilde x)
-
F_\succ(\ell;x^*)
\right|
\leq
\epsilon.
\]
Therefore there exists a realization of \(\widetilde x\) that is an
\(\epsilon\)-approximate solution and satisfies the top-\(k\) conclusion of the
theorem.

Moreover, for every \(\ell\in\{1,\ldots,n\}\),
\begin{align*}
\left|
T_\ell(\widetilde x)-T_\ell(x^*)
\right|
&=
\left|
\sum_{\succ\in\mcR}
\eta
\left[
F_\succ(\ell;\widetilde x)
-
F_\succ(\ell;x^*)
\right]
\right|
\\
&\leq
\sum_{\succ\in\mcR}
\eta
\left|
F_\succ(\ell;\widetilde x)
-
F_\succ(\ell;x^*)
\right| \leq \epsilon.
\end{align*}

Finally, by Lemma~\ref{lem:bvn}, each vertex of \(X\) has exactly one nonzero
entry in each row. Since
\[
\widetilde x
=
\frac{1}{K_\epsilon}
\sum_{i=1}^{K_\epsilon}
\widetilde m_i,
\]
each row of \(\widetilde x\) can therefore have at most \(K_\epsilon\) nonzero
entries.

For \(|\mcR|\geq2\),
\[
|\mcR|(|\mcR|-1)+1
\leq
|\mcR|^2.
\]
Therefore,
\[
K_\epsilon \leq 1+ \frac{2}{\epsilon^2}
\ln\left( |\mcR|^2 \right) = 1+ \frac{4\ln|\mcR|}{\epsilon^2}.
\]
Consequently, each represented preference type is assigned with positive
probability to at most $1+\frac{4\ln|\mcR|}{\epsilon^2}$ outcomes.
In particular, if \(|\mcR|\) is polynomial in \(n\), then each preference type is
assigned to at most $O\left( \frac{\ln n}{\epsilon^2} \right)$ outcomes.
\end{proof}

\newpage
\setcounter{section}{0}
\clearpage
\setcounter{page}{1} 
\appendix

\begin{center}
    \LARGE{\textbf{Online Appendix (not for publication)}}
\end{center}
\resetcounters

\section{Third case in the Proof of \autoref{2schools}}\label{OA_Proof2}

\begin{proof}
Consider the case when \(IC_B\) is violated at
\((\bar\epsilon,\bar\delta)\), or holds there with equality. Set
\[
\delta^*=\bar\delta
\]
and choose \(\epsilon^*\) so that \(IC_B\) binds:
\[
\bar\delta
=
\frac{u_2}{u_1}
\frac{\eta_A}{\eta_B}\epsilon^*
-
(1-c_A)(1-c_B)
\left(
1-\frac{u_2}{u_1}
\right)\eta_A.
\]
Equivalently,
\[
\epsilon^*
=
\left[
\bar\delta
+
(1-c_A)(1-c_B)
\left(
1-\frac{u_2}{u_1}
\right)\eta_A
\right]
\frac{u_1}{u_2}
\frac{\eta_B}{\eta_A}.
\]
Since \(IC_B\) is violated at the northeast corner, or binds there,
\[
\epsilon^*\leq\bar\epsilon.
\]

We verify the remaining cutoff and capacity restrictions. Since
\(\epsilon^*\leq\bar\epsilon\),
\[
\frac{\eta_A}{\eta_B}(1-c_B)\epsilon^*
\leq
\frac{\eta_A}{\eta_B}(1-c_B)\bar\epsilon
\leq
\bar\delta
\]
where we use \autoref{eq:bar_epsilon} for the last inequality.

Moreover, because \(IC_B\) binds and its right-hand side is strictly
below
\[
\frac{\eta_A}{\eta_B}\epsilon^*,
\]
we have
\[
\bar\delta
<
\frac{\eta_A}{\eta_B}\epsilon^*
<
\frac{\eta_A}{\eta_B}
\frac{\epsilon^*}{1-c_A}.
\]
Hence \((\epsilon^*,\bar\delta)\) satisfies all the cutoff and capacity restrictions. It also satisfies \(IC_A\) strictly.

Finally, for every feasible pair \((\epsilon,\delta)\),
\[
\delta\leq\bar\delta.
\]
Since \(IC_B\) requires
\[
\delta
\geq
\frac{u_2}{u_1}
\frac{\eta_A}{\eta_B}\epsilon
-
(1-c_A)(1-c_B)
\left(
1-\frac{u_2}{u_1}
\right)\eta_A,
\]
and its right-hand side is strictly increasing in \(\epsilon\), we
must also have
\[
\epsilon\leq\epsilon^*.
\]
Thus \((\epsilon^*,\bar\delta)\) weakly dominates every feasible pair coordinatewise and is therefore optimal. Hence
\[
\delta^*=\bar\delta,
\]
\(IC_B\) binds, and \(IC_A\) is slack. This is case~(iii).

\end{proof}

\section{Proofs}
\label{app:proof-stb-solution}

\subsection{Why the optimal STB adjustment leaves school $A$ unchanged}

We consider the planner's problem directly at the level of assignments.
Restrict attention to feasible incentive-compatible assignments that fill both schools.
Let \(y\) denote the mass of \(A\succ B\) students assigned to \(B\), and let \(z\) denote the mass of \(B\succ A\) students assigned to \(A\).
The matched part of any such assignment can then be written as
\[
\begin{array}{c|cc}
 & A & B \\ \hline
A\succ B & q_A-z & y \\
B\succ A & z & q_B-y .
\end{array}
\]
The total mass of first-choice assignments is
\[
F=q_A+q_B-y-z.
\]
Since both schools fill their capacities, the planner's payoff is
\[
\alpha F+(1-\alpha)(q_A+q_B-F),
\]
which, because \(\alpha>1/2\), is strictly decreasing in \(y+z\).
Thus the planner seeks to minimize the total mass of second-choice assignments.

The incentive constraint for students of type \(A\succ B\) is
\[
\frac{(q_A-z)u_1+yu_2}{\eta_A}
\geq
\frac{zu_1+(q_B-y)u_2}{\eta_B}.
\]
Using \(\eta_A+\eta_B=1\), this is equivalent to
\begin{equation}
y-\frac{u_1}{u_2}z
\geq
\eta_Aq_B-\eta_Bq_A\frac{u_1}{u_2}.
\label{eq:stb-reduced-ica}
\end{equation}
Similarly, incentive compatibility for students of type \(B\succ A\) requires
\begin{equation}
z-\frac{u_1}{u_2}y
\geq
\eta_Bq_A-\eta_Aq_B\frac{u_1}{u_2}.
\label{eq:stb-reduced-icb}
\end{equation}
Feasibility with respect to the masses of the two preference types requires
\begin{equation}
q_B-\eta_B
\leq
y-z
\leq
\eta_A-q_A.
\label{eq:stb-type-feasibility}
\end{equation}

We next show that assigning \(B\succ A\) students to \(A\) can never be optimal.
Consider any full-capacity incentive-compatible assignment with \(z>0\).
If \(y\geq z\), define $\widehat z=0$ and   $\widehat y=y-z$. By \eqref{eq:stb-type-feasibility}, the resulting assignment remains feasible.
Moreover,
\[
\widehat y
=
\left(
y-\frac{u_1}{u_2}z
\right)
+
\left(
\frac{u_1}{u_2}-1
\right)z,
\]
so \eqref{eq:stb-reduced-ica} remains satisfied, and
\[
-\frac{u_1}{u_2}\widehat y
=
\left(
z-\frac{u_1}{u_2}y
\right)
+
\left(
\frac{u_1}{u_2}-1
\right)z,
\]
so \eqref{eq:stb-reduced-icb} also remains satisfied.
The total mass of second-choice assignments falls from \(y+z\) to \(y-z\).

If instead \(y<z\), set $\widehat y=\widehat z=0.$ Since \eqref{eq:stb-type-feasibility} then implies \(q_B<\eta_B\), while \(c_A>0\) implies \(q_A<\eta_A\), the resulting assignment is feasible.
Moreover, \eqref{eq:stb-reduced-ica} at the original assignment implies $\eta_Aq_B-\eta_Bq_A\frac{u_1}{u_2}<0$, while $\frac{\eta_A}{q_A}>\frac{\eta_B}{q_B}$ and \(u_1>u_2\) imply $\eta_Bq_A-\eta_Aq_B\frac{u_1}{u_2}<0$. Hence both incentive constraints are satisfied at \((\widehat y,\widehat z)=(0,0)\). Again, the mass of second-choice assignments is strictly reduced.

It follows that any optimal full-capacity incentive-compatible assignment must have $z=0$. Thus all seats at school \(A\) are optimally assigned to students who rank \(A\) first.

\subsection{Proof of Proposition~\ref{prop:stb-solution}}

\begin{proof}
The assumptions imply that $1>c_A>c_B>0$. Preserving capacity at $B$ requires $\eta_A\epsilon=\eta_B\epsilon'$, so $\epsilon'=(\eta_A/\eta_B)\epsilon$.

A student who ranks $A$ first obtains expected utility $(1-c_A)u_1+(c_A-c_B-\epsilon)u_2$ from truthful reporting and $(1-c_B+\epsilon')u_2$ from reporting $B\succ A$. Incentive compatibility therefore requires
\[
(1-c_A)u_1+(c_A-c_B-\epsilon)u_2 \geq(1-c_B+\epsilon')u_2.
\]
Using $\epsilon'=(\eta_A/\eta_B)\epsilon$ and $\eta_A+\eta_B=1$, this becomes $\epsilon\leq(1-c_A)\eta_B(u_1/u_2-1)$.

For a student who ranks $B$ first, the gain from truthful reporting over reporting $A\succ B$ is $(1-c_A)(u_1-u_2)+(\epsilon+\epsilon')u_1>0$. Thus their incentive constraint is strictly satisfied.

The cutoff restrictions require $c_B+\epsilon\leq c_A$, or $\epsilon\leq c_A-c_B$, and $c_B-\epsilon'\geq0$, or $\epsilon\leq(\eta_B/\eta_A)c_B$.

The change in the planner's payoff is $(2\alpha-1)\eta_A\epsilon$, which is strictly increasing in $\epsilon$. The optimal adjustment therefore chooses the largest feasible value,
\[
\epsilon^*
=
\min\left\{
(1-c_A)\eta_B\left(\frac{u_1}{u_2}-1\right),
\;c_A-c_B,
\;\frac{\eta_B}{\eta_A}c_B
\right\},
\]
with $\epsilon^{\prime *}=(\eta_A/\eta_B)\epsilon^*$.
\end{proof}

\section{The cutoff construction under single tie-breaking}
\label{example:imple-STB}

Consider the assignment in Example~\ref{example:imple-MTB}, which gives lotteries $p_{\succ_A}=(3/5,1/5,1/5)$ and $p_{\succ_B}=(1/5,3/5,1/5)$, with coordinates corresponding to $(A,B,\emptyset)$. Each student receives their first choice with probability $3/5$ and, conditional on rejection from their first choice, receives their second choice with probability $1/2$.

Under STB, each student draws a single lottery number $\omega\sim U[0,1]$, which determines priority at both schools. As in Example~\ref{example:imple-MTB}, let $c_A$ and $c_B$ be the school cutoffs, and let $\rho_{AA},\rho_{AB},\rho_{BA},\rho_{BB}$ be the report-dependent bonuses. A student reporting $A\succ B$ has scores $\omega+\rho_{AA}$ at $A$ and $\omega+\rho_{AB}$ at $B$, and similarly for a student reporting $B\succ A$.

Only the difference between a cutoff and a bonus matters for eligibility, so we normalize the first-choice bonuses to zero: $\rho_{AA}=\rho_{BB}=0$. A student reporting $A\succ B$ must receive $A$ with probability $3/5$, so $\Pr(\omega\geq c_A)=3/5$, which gives $c_A=2/5$. The same argument for students reporting $B\succ A$ gives $c_B=2/5$.

Conditional on rejection from $A$, a student reporting $A\succ B$ must be eligible for $B$ with probability $1/2$. Thus
\[
\Pr\left(
\omega\geq\frac25-\rho_{AB}
\,\middle|\,
\omega<\frac25
\right)=\frac12.
\]
Conditional on $\omega<2/5$, the lottery number is uniform on $[0,2/5)$, so $\rho_{AB}=1/5$. By symmetry, $\rho_{BA}=1/5$.

Thus the assignment is reproduced by setting $c_A=c_B=2/5$, giving no bonus to students who rank a school first, and giving a bonus of $1/5$ to students who rank it second.

\end{document}